\documentclass[aps,prx,preprint,onecolumn,citeautoscript,superscriptaddress,footinbib,
eqsecnum]{revtex4-1}  
\usepackage{amsmath,amssymb,bm,bbm} 
\usepackage{graphicx}  
\usepackage{color} 
\usepackage[dvipsnames]{xcolor}
\usepackage[papersize={8.5in,11in}]{geometry}
\usepackage[colorlinks=true]{hyperref}  
\hypersetup{
    bookmarks=true,         
    unicode=false,          
    pdftoolbar=true,        
    pdfmenubar=true,        
    pdffitwindow=false,     
    pdfstartview={FitH},    
    pdftitle={Fermionic spinon theory of square lattice spin liquids near the Neel state},    
    pdfauthor={Alex Thomson and Subir Sachdev},     
    pdfsubject={},   
    pdfcreator={},   
    pdfproducer={}, 
    pdfkeywords={} {} {}, 
    pdfnewwindow=true,      
    colorlinks=true,       
    linkcolor=magenta, 
    citecolor=blue,        
    filecolor=magenta,      
    urlcolor=blue           
} 

\usepackage{amsfonts}
\usepackage{upgreek}
\usepackage{slashed}
\usepackage{latexsym}
\usepackage{todonotes}

\newcommand{\beq}{\begin{equation}}
\newcommand{\eeq}{\end{equation}}
\def\bea{\begin{eqnarray}}
\def\eea{\end{eqnarray}}

\def \Q{{\bm Q}}

\newcommand{\nn}{\nonumber \\}

\usepackage{braket}
\usepackage{enumitem}
\usepackage{math tools} 
\usepackage{dsfont} 
\usepackage{mathrsfs}
\usepackage{multirow}

\usepackage{subcaption}
\usepackage{pifont}
\usepackage{colortbl}
\usepackage{hhline} 
\makeatletter
\def\hlinewd#1{%
\noalign{\ifnum0=`}\fi\hrule \@height #1 %
\futurelet\reserved@a\@xhline}
\makeatother

\makeatletter

    \def\CT@@do@color{%
      \global\let\CT@do@color\relax
            \@tempdima\wd\z@
            \advance\@tempdima\@tempdimb
            \advance\@tempdima\@tempdimc
    \advance\@tempdimb\tabcolsep
    \advance\@tempdimc\tabcolsep
    \advance\@tempdima2\tabcolsep
            \kern-\@tempdimb
            \leaders\vrule
                    \hskip\@tempdima\@plus  1fill
            \kern-\@tempdimc
            \hskip-\wd\z@ \@plus -1fill }
    \makeatother

\newcommand{\gW}{{\mathfrak{W}}}
\newcommand{\bch}{\bar{\chi}}
\newcommand{\bX}{\bar{X}}

\renewcommand{\O}{\mathcal{O}}

\newcommand{\M}{\mathcal{M}}

\newcommand{\T}{\mathcal{T}}

\renewcommand{\L}{\mathcal{L}}

\newcommand{\cX}{{\mathcal{X}}}

\newcommand{\G}{{\mathscr{G}}}
\newcommand{\sT}{{\mathscr{T}}}

\newcommand{\vk}{{\boldsymbol{k}}}
\newcommand{\br}{{\boldsymbol{r}}}

\newcommand{\vi}{{\boldsymbol{i}}}
\newcommand{\vj}{{\boldsymbol{j}}}
\newcommand{\vm}{{\boldsymbol{m}}}

\newcommand{\hx}{{\hat{\boldsymbol{x}}}}
\newcommand{\hy}{{\hat{\boldsymbol{y}}}}

\newcommand{\vphi}{{\varphi}}

\newcommand{\bpsi}{\bar{\psi}}

\newcommand{\tr}{\text{tr}}
\newcommand{\sd}[1]{\slashed{#1}}

\renewcommand{\o}{\over}
\newcommand{\eq}[1]{\begin{align}#1\end{align}}

\newcommand{\diag}{\mathrm{diag}}

\renewcommand{\(}{\left(}
\renewcommand{\)}{\right)}
\renewcommand{\[}{\left[}
\renewcommand{\]}{\right]}
\newcommand{\abs}[1]{\left| #1 \right|}

\newcommand{\nt}{\notag\\}

\let\v\boldsymbol

\newcommand{\ph}{\phantom}

\newcommand{\ep}{\epsilon}
\newcommand{\vep}{\varepsilon}
\renewcommand{\a}{\alpha}
\renewcommand{\b}{\beta}
\renewcommand{\d}{\delta}
\newcommand{\g}{\gamma}
\newcommand{\n}{\nu}
\newcommand{\m}{\mu}
\renewcommand{\t}{\tau}
\newcommand{\s}{\sigma}

\renewcommand{\th}{\theta}

\newcommand{\lam}{\lambda}

\let\ptl\partial
\renewcommand{\dag}{\dagger}

\newcommand{\id}{\mathds{1}}
\newcommand{\Zt}{\mathds{Z}_2}

\renewcommand\Re{\operatorname{\mathfrak{Re}}}
\renewcommand\Im{\operatorname{\mathfrak{Im}}}

\newcommand{\ua}{\uparrow}
\newcommand{\da}{\downarrow}

\newcommand{\biz}{\begin{itemize}}
\newcommand{\eiz}{\end{itemize}}

\newcommand{\qcd}{{\mathrm{QCD}_3}}

\begin{document}

\title{Fermionic spinon theory of square lattice spin liquids\\ near the N\'eel state}

\author{Alex Thomson}
\affiliation{Department of Physics, Harvard University, Cambridge MA 02138, USA}
\affiliation{Kavli Institute for Theoretical Physics, University of California, Santa Barbara, California 93106, USA}

\author{Subir Sachdev}
\affiliation{Department of Physics, Harvard University, Cambridge MA 02138, USA}
\affiliation{Perimeter Institute for Theoretical Physics, Waterloo, Ontario, Canada N2L 2Y5}

\date{\today
\\
\vspace{0.4in}}

\begin{abstract}
Quantum fluctuations of the N\'eel state of the square lattice
antiferromagnet are usually described by a $\mathbb{CP}^1$
theory of bosonic spinons coupled to a U(1) gauge field, and with a global SU(2) spin rotation symmetry.
Such a theory also has a confining phase with valence bond solid (VBS) order, and upon including spin-singlet charge 2 Higgs fields, deconfined phases
with $\mathbb{Z}_2$ topological order possibly intertwined with discrete broken global symmetries. We present dual theories of the same phases starting from a mean-field theory of fermionic spinons 
moving in $\pi$-flux in each square lattice plaquette. Fluctuations
about this $\pi$-flux state are described by 2+1 dimensional
quantum chromodynamics (QCD$_3$) with a SU(2) gauge group and $N_f=2$ flavors of massless Dirac fermions. It has recently been argued by Wang {\em et al.\/} (arXiv:1703.02426) that this QCD$_3$ theory describes the N\'eel-VBS quantum phase transition. We introduce adjoint Higgs fields in QCD$_3$, and obtain fermionic dual descriptions of the phases with $\mathbb{Z}_2$ topological order obtained earlier using the bosonic $\mathbb{CP}^1$ theory. We also present a fermionic spinon derivation of the monopole Berry phases in the U(1) gauge theory of the VBS state. The global phase diagram of these phases contains multi-critical points, and our results imply new boson-fermion dualities between critical gauge theories of these points.

\end{abstract}
\maketitle

\section{Introduction}
\label{sec:intro}

Spin liquid states of the square lattice antiferromagnet, with global SU(2) spin rotation symmetry, have long been recognized as important ingredients in the theory of the cuprate high temperatures superconductors \cite{PWA87,GBPWA88,KRS88,DRSK88}. The earliest established examples of gapped states were `chiral spin liquids,' which were constructed by analogy to the fractional quantum Hall states \cite{Kalmeyer87,WWZ89}. These have a topological order which is not compatible with time-reversal symmetry. Soon after, `$\mathbb{Z}_2$ spin liquids' were proposed \cite{NRSS91,SSNR91,XGW91,Bais92,SenthilFisher,Freedman04,Hansson04}: their topological order is compatible with time-reversal symmetry, and exactly solvable examples were later found in Kitaev's toric code and honeycomb lattice models \cite{Kitaev03,Wen03,Kitaev06}. Wen \cite{Wen02} used a fermionic spinon representation of the antiferromagnet to obtain a plethora of possible square lattice spin liquid states, distinguished by different realizations of `symmetry-enriched' topological order \cite{Wen11,EH13}. Wen's classification criterion was that the spin liquid states preserve time-reversal, SU(2) spin rotations, and all the square lattice space group symmetries. However, in the application to the cuprates, there is no fundamental reason all such symmetries should be preserved. If we also allow for breaking of time-reversal and/or point group symmetries, then many more spin liquid states are clearly possible, all of which preserve SU(2) spin rotations and the square lattice translational symmetry \cite{NRSS91,SSNR91,BYK12,CQSS16}. This proliferation of possible spin liquids, intertwining with broken symmetries, sets up a daunting task of deciding which states, if any, are relevant for the pseudogap phase of the underdoped cuprates.

We need an energetic and physical criterion to focus on a smaller set of relevant spin liquid states, rather
than relying exclusively on symmetry and topology. 
In recent work, Chatterjee {\em et al.\/} \cite{CSS17} proposed examining spin liquids which are proximate to the magnetically ordered N\'eel state. These proximate states are reachable by continuous (or nearly continuous) quantum phase transitions involving the long-wavelength excitations of the antiferromagnet. Specifically, they used a $\mathbb{CP}^1$ theory of quantum fluctuations of the N\'eel state, expressed in terms of bosonic spinons, $z_\alpha$, to argue for the importance of 3 possible $\mathbb{Z}_2$ spin liquid states. These 3 states are identified here as $A_b$, $B_b$, and $C_b$, and appear below in Figs.~\ref{fig:1}a and~\ref{fig:2}a. The state $A_b$ preserves all symmetries \cite{YangWang}, while $B_b$ breaks lattice rotation symmetries and so has Ising-nematic order \cite{NRSS91}. The state $C_b$ breaks inversion and time-reversal symmetries, but not their product, and was argued to possess current loop order.

A related motivation for the physical importance of these states comes from an examination of the classical phase diagram of frustrated antiferromagnets on the square lattice. By examining models with two-spin near-neighbor and four-spin ring exchange interactions, Ref.~\onlinecite{CSS17}
found magnetically ordered states with canted, spiral, and conical spiral order near the N\'eel state. 
Quantum fluctuations about these classical ordered states can be described by extensions of the $\mathbb{CP}^1$ theory, and the `quantum disordered'
states obtained across a continous transition involving loss of magnetic order are
precisely the three $\mathbb{Z}_2$ spin liquids, with the correspondence \cite{CS17,CSS17}
\beq
    \mbox{canted order}  \rightarrow A_b\,,\quad
     \mbox{spiral order}  \rightarrow B_b\,,\quad
      \mbox{conical spiral order} \rightarrow C_b\,.
\eeq

One of the purposes of the present paper is to present a unified theory of the 3 $\mathbb{Z}_2$ spin liquids noted above, but using the fermionic spinon approach \cite{AM88,Affleck88,Wen02}. For gapped $\mathbb{Z}_2$ spin liquid states, a mapping between the fermionic and bosonic spinons approaches has been achieved for specific states on the kagome, triangular, square, and rectangular lattices \cite{YY12,EH13,LCV14,QiFu15,ZLV15,ZMQ15,Lu15,YangWang,CQSS16}. This mapping relies on the fusion rules of the toric code \cite{Kitaev03}: the fusion of any two of the anyon species yields the third. In $\mathbb{Z}_2$ spin liquids, the three types of anyons are bosonic spinons, fermionic spinons, and a bosonic $\mathbb{Z}_2$-flux spinless vison. We will extend such mappings here to the states of interest on the square lattice, but using a method which allows us to treat the 3 $\mathbb{Z}_2$ spin liquids and the quantum phase transitions between them in a unified manner.  We will obtain a phase diagram of the states proximate to the N\'eel state using the fermionic spinon approach, and propose critical theories of the phase transitions involving massless Dirac fermions. The connection to the earlier analysis \cite{CSS17} using the bosonic spinons of the $\mathbb{CP}^1$ model will also lead us to propose new boson-fermion dualities of the strongly-coupled, gapless, quantum field theories describing the (multi-)critical points. 

Our point of departure will be a boson-fermion duality of a conformal field theory (CFT) proposed by Wang {\em et al.\/} \cite{Wang17}. They examined the critical theory of the N\'eel-valence bond solid (VBS) transition in the $\mathbb{CP}^1$ theory \cite{NRSS89,NRSS90,senthil1,senthil2}, and proposed that it was equivalent to quantum chromodynamics (QCD) with a SU(2) gauge group and $N_f=2$ flavors of massless, two-component Dirac fermions (note: the SU(2) gauge group is not to be confused with the global SU(2) spin rotation symmetry). The latter theory can also be obtained from the fermionic spinon approach to the square lattice antiferromagnet: it describes fluctuations about a $\pi$-flux mean-field theory \cite{AM88,Affleck88,Wen02}, which is labeled by Wen as SU2B$n0n1$.

Starting from the SU(2) QCD$_3$ theory, we will explore routes to condensing Higgs fields for fermionic bilinears, so that the SU(2) gauge group is ultimately broken down to $\mathbb{Z}_2$ and we obtain gapped spin liquids with $\mathbb{Z}_2$ topological order. Our main results are contained in the phase diagrams in Fig.~\ref{fig:2}. These phase diagrams contain the phases $A_f$, $B_f$ and $C_f$, which are fermionic counterparts of the $A_b$, $B_b$, and $C_b$ states obtained from the bosonic $\mathbb{CP}^1$ theory.

One important feature of the fermionic phase diagram in Fig.~\ref{fig:1}b is that it does not contain the counterparts of the magnetically ordered N\'eel and canted states in the bosonic phase diagram in  Fig.~\ref{fig:1}a. Instead Fig.~\ref{fig:1}b contains two critical phases, with massless Dirac fermions interacting with gapless SU(2) and U(1) gauge bosons. Building on the fermion-boson equivalence of Wang {\em et al.\/} \cite{Wang17}, we argue here that these critical phases of Fig.~\ref{fig:1}b are unstable to the corresponding magnetically ordered phases in Fig.~\ref{fig:1}a; the instability is assumed to be driven by relevant operators which are allowed by the symmetries of the underlying square lattice antiferromagnet. However, given the strongly-coupled nature of the critical theories, this conclusion is based upon circumstantial, rather than firm, evidence.

\subsection{Summary of results}

\begin{figure}
	\centering
	\includegraphics[width=7in]{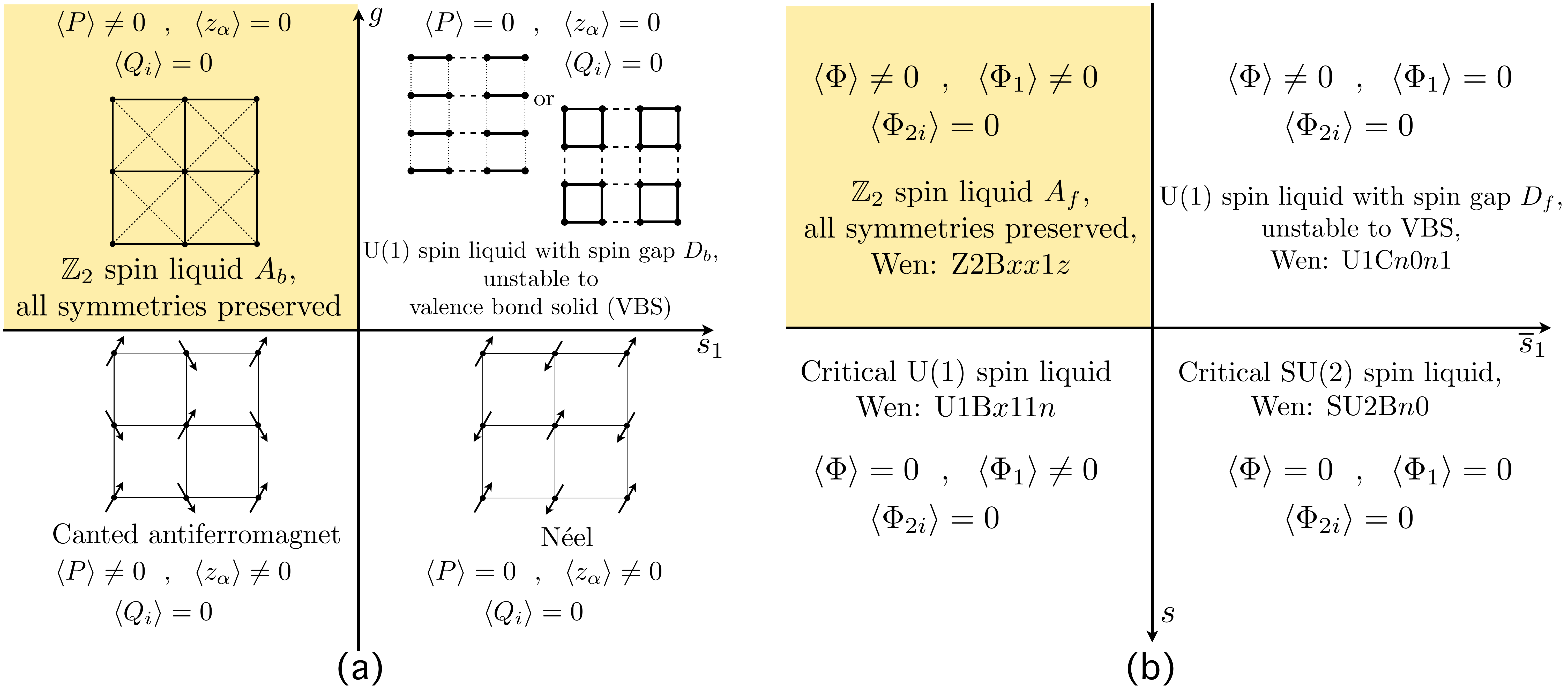}
	\caption{(a) Schematic phase diagram of the $\mathbb{CP}^1$ theory in Eq.~(\ref{b1}) as a function of 
	$g$ and $s_1$ ($s_2$ in Eq.~(\ref{b2}) is large and positive); Eq.~(\ref{dcp}) describes the deconfined critical N\'eel-VBS transition at a critical $g=g_c$. 
	(b) Schematic phase diagram of the SU(2) QCD$_3$ theory with $N_f=2$ flavors of massless Dirac fermions 
	in Eq.~(\ref{f2}) as a function of 
	$s$ and $\overline{s}_1$ ($\overline{s}_2$ in Eq.~(\ref{f3}) is large and positive). The `Wen' labels refer to the naming scheme in Ref.~\onlinecite{Wen02}. The $\mathbb{Z}_2$ spin liquids $A_b$ and $A_f$ in (a) and (b) are argued to be topologically identical, as are the confining states with VBS order.
	The critical spin liquids in (b) to be unstable
	to the corresponding phases with magnetic order in (a), with the critical SU(2) spin liquid surviving only at the N\'eel-VBS transition. All $\mathbb{Z}_2$ spin liquids are shown shaded in all figures.}
	\label{fig:1}
\end{figure}
Let us first recall the bosonic spinon approach \cite{AA89,CSS17} to the phases in Fig.~\ref{fig:1}a. 
This is obtained by extending the Lagrangian for the 
theory of deconfined criticality for the N\'eel-VBS transition \cite{SJ90}
\beq
\mathcal{L}_{dcp} = \frac{1}{g} \left| (\partial_\mu - i b_\mu) z_\alpha \right|^2 + S_B \,.
\label{dcp}
\eeq
The Lagrangian is in three spacetime dimensions with
$\mu$ a spacetime index in Minkowski signature $(+,-,-)$,
and $\alpha,\beta = \uparrow,\downarrow$ so there is global SU(2) spin rotation symmetry. 
The N\'eel order parameter is $z_\alpha^\ast \sigma^a_{\alpha\beta} z_\beta$, where $\sigma^a$ are the Pauli matrices.
The U(1) gauge field $b_\mu$ is compact,
and monopole tunneling events are permitted, and associated with a Berry phase $S_B$ \cite{Haldane88,NRSS90}. The spinons are represented by the bosonic complex scalar $z_\alpha$ which 
is of unit length
\begin{equation}
    \sum_\alpha |z_\alpha|^2 = 1\,,
\end{equation}
and carries unit U(1) charge. For small $g$, $z_\alpha$ is condensed, and this yields the N\'eel phase with broken spin rotation symmetry. For large $g$, $z_\alpha$ is not condensed, 
and we appear to obtain a U(1) spin liquid
(which we call $D_b$) with a gapless photon $b_\mu$, and gapped $z_\alpha$ spinons. However, the condensation of monopoles yields the confinement of spinons and the appearance of VBS order \cite{NRSS89,NRSS90}. 
The transition from the N\'eel state to the VBS is described by a deconfined critical theory \cite{senthil1,senthil2} at $g=g_c$ in which monopoles are suppressed.

We will now extend $\mathcal{L}_{dcp}$ by including  complex, charge 2 Higgs fields whose condensation can induce phases with $\mathbb{Z}_2$ topological 
order, while preserving SU(2) spin rotation symmetry. 
We can construct such Higgs fields by pairing spinons, but the simplest possibility, $\varepsilon_{\alpha\beta} z_\alpha z_\beta$, vanishes identically. Any such spinon pair Higgs field must involve gradients, and the simplest non-vanishing cases involve a single temporal or spatial gradient. We consider first the Higgs field, $P$, conjugate to a pair of spinons with a single temporal gradient, and will include the spatial gradient Higgs field, $Q_i$ later.
The Lagrangian for $z_\alpha$, $b_\mu$, and $P$ is 
\beq
\mathcal{L}_b = \mathcal{L}_{dcp} + |(\partial_\mu - 2 i b_\mu) P|^2 - s_1 |P|^2 +  \lambda_1 \, P^\ast \, \varepsilon_{\alpha\beta} z_\alpha
\partial_0 z_\beta +  \lambda_1 \, P \, \varepsilon_{\alpha\beta} z_\alpha^\ast
\partial_0 z_\beta^\ast + \ldots\,,
\label{b1}
\eeq
where $\varepsilon_{\alpha\beta}$ is the unit anti-symmetric tensor, and so SU(2) spin rotation symmetry is maintained. 
For $s_1$ large and positive, when there is no $P$ condensate, we obtain the phases of $\mathcal{L}_{dcp}$
already described. For smaller $s_1$, when there is a $P$ Higgs condensate, we obtain
the canted antiferromagnet and the symmetric $\mathbb{Z}_2$ spin liquid $A_b$ for small and large $g$ respectively,
as shown in Fig.~\ref{fig:1}a. The $\mathbb{Z}_2$ spin liquid $A_b$ was first obtained in Ref.~\onlinecite{YangWang}, where it was called $\mathbb{Z}_2[0,0]$.

Now we turn to our results for the fermionic counterpart of Fig.~\ref{fig:1}a, which is shown in Fig.~\ref{fig:1}b.
We start with fermionic equivalent of the deconfined N\'eel-VBS critical theory, which was identified
by Wang {\em et al.\/} \cite{Wang17} as SU(2) QCD$_3$ with $N_f=2$, described by the Lagrangian
\begin{equation}
    \mathcal{L}_{\mathrm{QCD}_3} = i \mbox{tr} \left( \bar{X} \gamma^\mu (\partial_\mu X + i X a_\mu) \right)
    \label{f1}
\end{equation}
Here $X$ represents the massless Dirac fermions, $\gamma^\mu$ are Dirac matrices, and the details of the index structure will be specified in Section~\ref{sec:PiFluxAnsatz}. 
The SU(2) gauge field is represented by $a_\mu$. The fermion kinetic term in Eq.~(\ref{f1}) has
a global SO(5) symmetry, which is an enlargement of the global SU(2) spin rotation and $\mathbb{Z}_4$ lattice rotation symmetries of the lattice Hamiltonian \cite{Wang17}.
To obtain Fig.~\ref{fig:1}b, we extend Eq.~(\ref{f1}) in Section~\ref{sec:symmliq}
by adding two real Higgs fields, $\Phi=\Phi^a\sigma^a$ and 
$\Phi_1=\Phi_1^a\sigma^a$, both of which transform as adjoints of the gauge SU(2). So we have the Lagrangian
\bea \label{eqn:HiggsLagrangian1}
    \mathcal{L}_f = \mathcal{L}_{\mathrm{QCD}_3} &+& (D_\mu \Phi^a)^2 - s \left(\Phi^a\right)^2 + \lambda_2 \, \Phi^a 
   \,  \mbox{tr} \left( \sigma^a \bar{X} \mu^y X \right) \nn
   &+& (D_\mu \Phi_1^a)^2 - \overline{s}_1 \left(\Phi_1^a\right)^2 + i\lambda_3 \, \Phi_1^a 
   \,  \mbox{tr} \left( \sigma^a \bar{X} \partial_0 X \right) + \ldots
   \label{f2}
\eea
Here $D_\mu$ is a covariant derivative, $a$ is SU(2) gauge index, $\sigma^a$ are Pauli matrices, while $\mu^y$ is a Pauli matrix which
acts on the flavor space. We assume the higher order terms are such that when both Higgs condensates
are present, $\langle \Phi \rangle$ and $\langle \Phi_1 \rangle$ will be oriented perpendicular to each other in SU(2) gauge space.
For instance, the topological order would be stabilized by the presence of a term like $-\m \(\Phi^a \Phi_1^a \)^2$ when $\m>0$.
By varying $s$ and $\overline{s}_1$ we can obtain four phases in which the two
Higgs condensates are either present or absent, as shown in Fig.~\ref{fig:1}b. We will show in Section~\ref{sec:PSGcorres} that
the gapped $\mathbb{Z}_2$ spin liquid, $A_f$, so obtained is topologically identical to the $\mathbb{Z}_2$ spin liquid
$A_b$ in Fig.~\ref{fig:1}a. 

We will also examine the U(1) spin liquid with a spin gap, $D_f$, obtained when there is only a $\Phi$ condensate.
We compute the monopole Berry phases in this state in Section~\ref{sec:BerryPhase}, and find that they are identical to those indicated
by $S_B$ in the bosonic theory in Eq.~(\ref{b1}). As monopoles are eventually expected to proliferate
in this U(1) spin liquid \cite{NRSS89}, we expect VBS order to appear, just as in the corresponding
phase in Fig.~\ref{fig:1}a.

Now we turn our attention to the critical U(1) and SU(2) spin liquids in Fig.~\ref{fig:1}b. As we noted
earlier, we expect that in the absence of fine-tuning, there are relevant perturbations to $\mathcal{L}_f$
which will drive these critical phases to the corresponding magnetically ordered phases in Fig.~\ref{fig:1}a. These perturbations will break the SO(5) flavor symmetry of $\mathcal{L}_{\mathrm{QCD}_3}$ down
to the symmetries of the underlying lattice Hamiltonian \cite{Wang17}.

Finally, we note that both Figs.~\ref{fig:1}a and~\ref{fig:1}b contain multicritical points accessed
by tuning 2 couplings where all 4 phases meet. A natural conjecture is that these multicritical points
are identical to each other. On the bosonic side, this is the theory obtained by tuning $g$ and $s_1$,
so that both the matter fields $z_\alpha$ and $P$ are critical. On the fermionic side, this is
the theory obtained by tuning $s$ and $\overline{s}_1$, so that the bosonic
matter fields $\Phi$ and $\Phi_1$ are critical, while the fermionic matter $X$ remains critical. A further conjecture is that the Yukawa couplings $\lambda_1$
and $\lambda_3$ renormalize to zero at the multicritical point: then both the bosonic and fermionic
theories will represent CFTs.

\begin{figure}[tb]
	\centering
	\includegraphics[width=7in]{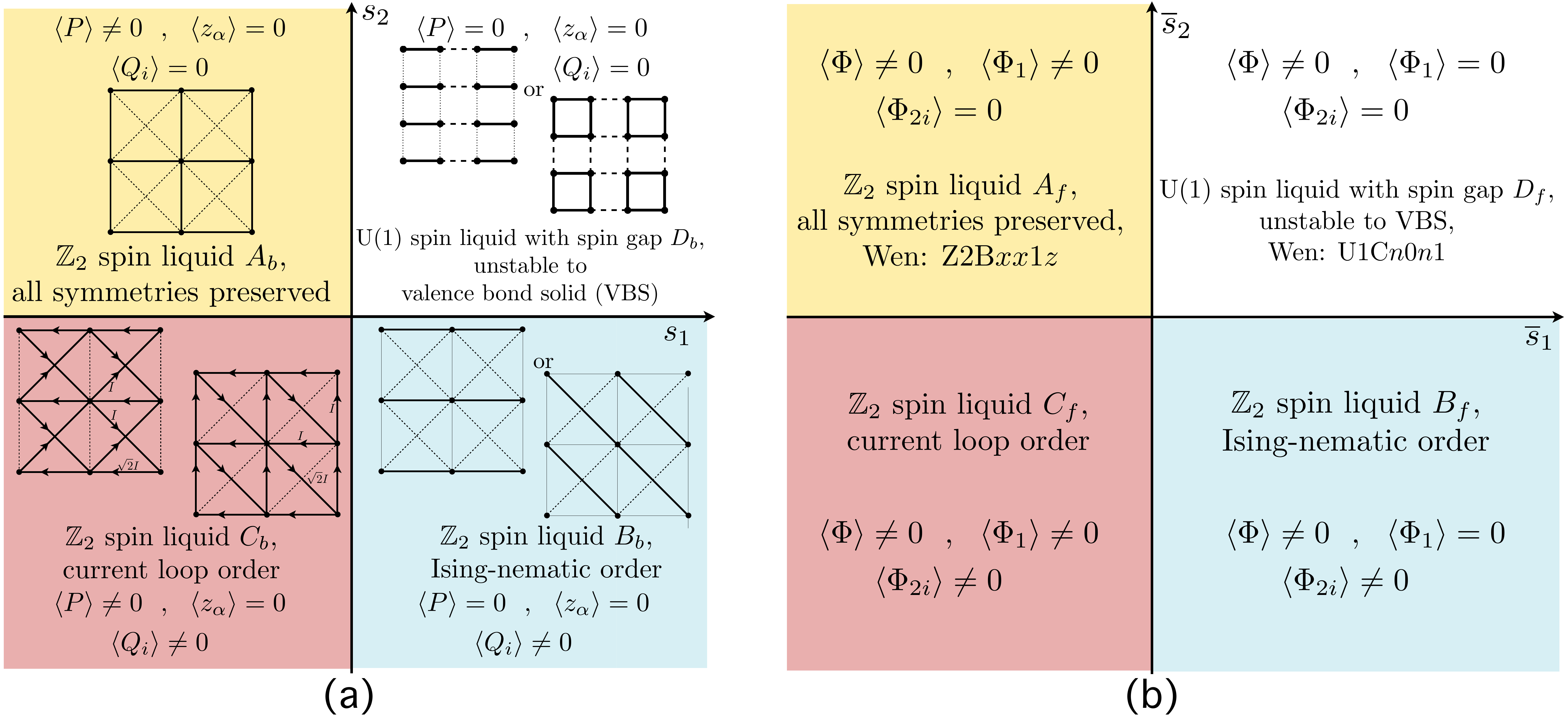}
	\caption{(a) Schematic phase diagram of the $\mathbb{CP}^1$ theory in Eqs.~(\ref{b1}) and (\ref{b2}) as a function of 
	$s_1$ and $s_2$ (for large $g$).
	(b) Schematic phase diagram of the SU(2) QCD$_3$ theory with $N_f=2$ flavors of massless Dirac fermions 
	in Eqs.~(\ref{f2}) and (\ref{f3}) as a function of 
	$\overline{s}_1$ and $\overline{s}_2$ (for $s<0$  and $|s|$ large). All four phases in (a) and (b) 
	are argued to be topologically identical. So for the $\mathbb{Z}_2$ spin liquids
	$A_b = A_f$, $B_b=B_f$, and $C_b=C_f$. Phases $B_f$ and $C_f$ do not appear in Wen's classification \cite{Wen02} because
	they break global symmetries.}
	\label{fig:2}
\end{figure}
We also extend our results to include additional Higgs fields which lead to phases with $\mathbb{Z}_2$
topological order and broken lattice rotation and/or time-reversal symmetries. 
On the bosonic side, we introduce the complex, charge 2 Higgs field $Q_i$, where $i=x,y$ is a spatial index,
leading to the Lagrangian \cite{SSNR91}
\begin{equation}
\mathcal{L}_b^\prime = \mathcal{L}_b + 
    |(\partial_\mu - 2 i b_\mu) Q_i|^2 - s_2 |Q_i|^2 +  \lambda_4 \, Q_i^\ast \, \varepsilon_{\alpha\beta} z_\alpha
\partial_i z_\beta +  \lambda_4 \, Q_i \, \varepsilon_{\alpha\beta} z_\alpha^\ast
\partial_i z_\beta^\ast + \ldots
\label{b2}
\end{equation}
In the absence of magnetic order, so that $g$ is large, the phase diagram obtained by varying
$s_1$ and $s_2$, with possible condensates of $P$ and $Q_i$ is shown in Fig.~\ref{fig:2}a.
There are now 3 $\mathbb{Z}_2$ spin liquids, and these meet at a possible multicritical point
with the VBS state.

On the fermionic side, in Section~\ref{sec:Isingnematic}, we add another real Higgs field, $\Phi_{2i}$, which transforms as the adjoint of SU(2). We now extend $\mathcal{L}_f$ in Eq.~(\ref{f2}) to 
\bea
    \mathcal{L}_f^\prime = \mathcal{L}_f + (D_\mu \Phi_{2i}^a)^2 - \overline{s}_2 \left(\Phi_{2i}^a\right)^2 + i\lambda_5 \, \Phi_{2i}^a 
   \,  \mbox{tr} \left( \sigma^a \bar{X} \partial_i X \right) + 
   \ldots
   \label{f3}
\eea
The phase diagram obtained by varying $\overline{s}_1$ and $\overline{s}_2$, to obtain possible
Higgs condensates of $\Phi_1$ and $\Phi_{2i}$ is shown in Fig.~\ref{fig:2}b.
We assume that $s$ is negative, so that a Higgs condensate $\Phi$ is always present in Fig.~\ref{fig:2}b. We obtain 3 $\mathbb{Z}_2$ spin liquids in Fig.~\ref{fig:2}b, and one of our
main results is that these are topologically identical to the corresponding $\mathbb{Z}_2$ spin liquids in Fig.~\ref{fig:2}a.
The relative orientations of the condensates of $\Phi$,
$\Phi_1$, and $\Phi_2$ in gauge space are discussed in Section~\ref{sec:currentloop}. Note that the spin liquids $B_f$ and $C_f$ do not appear in Wen's classification: this is because they break global symmetries associated with the appearance of Ising-nematic and current loop order respectively.

Again, the multicritical points in Figs.~\ref{fig:2}a and~\ref{fig:2}b, if present, are expected to map
to each other, setting up possible dualities of critical fermionic and bosonic gauge theories.

The paper is organized as follows.
In Sec.~\ref{sec:model}, we provide the background information necessary for our analysis.
We begin by discussing the relevant symmetries and reviewing the $\pi$-flux phase, showing that its low energy dynamics are described by $N_f=2$ QCD.
The section finishes with a brief summary of the boson-fermion duality proposed by Wang {\it et al.} \cite{Wang17}.
Sec.~\ref{sec:ProximateSpinLiquids} explains our procedure for finding spin liquids and how these phases are classified.
Using this, we next list all gapped spin liquids accessible using our methods and which are either fully symmetric or have Ising-nematic order. 
We also describe how spin liquids breaking additional discrete symmetries can be realized, with particular focus given to the $\Zt$ spin liquid $C_f$ with current-loop order.
These spin liquids, both symmetric and ordered, are subsequently identified in Sec.~\ref{sec:SpinLiqID}.
We start by using the symmetry fractionalization technique to verify the correspondence between the $\Zt$ spin liquids we study and those realized using Schwinger bosons.
This allows us to verify the equivalence of $A_f$, $B_f$, and $C_f$ with $A_b$, $B_b$, and $C_b$.
A comparison with Wen's \cite{Wen02} lattice classification scheme is also provided before we turn to the unstable U(1) spin liquid $D_f$ and demonstrate that the proliferation of monopoles necessarily results in a confined phase with VBS order.
We conclude in Sec.~\ref{sec:conc} with some discussion.

We note a related paper \cite{CXu17} which appeared 
while our work was being completed, describing phases of
antiferromagnets with only a U(1), `easy-plane', global spin rotation symmetry.

\section{$\pi$-flux phase and $N_f=2$ QCD}\label{sec:model}

\subsection{Model and symmetries}\label{sec:ModSyms}
We are interested in this paper in spin liquid states of  
the spin-1/2 Heisenberg model on the square lattice, with Hamiltonian of the form
\eq{
H_\mathrm{H}=J\sum_{\Braket{\vi\vj}} \v{S}_\vi\cdot \v{S}_\vj+\cdots\,,
} 
where the summation is over nearest-neighbours and the ellipsis indicates interactions over further distances or terms which comprise of three or more spin operators.
In the absence of these higher order terms, the ground state is known to have N\'{e}el order; nonetheless, we will operate under the assumption that the terms contained in the ellipsis provide enough frustration that the ground state loses long-range magnetic order.

It has been shown that a fully symmetric phase describing spin 1/2's on a square lattice must have topological order \cite{Lieb61,Hastings04}. 
It turns out that there are many possible such symmetric spin liquids, and a large body of work has been directed at classifying these phases.
One such scheme is provided by Wen in Ref.~\onlinecite{Wen02}. 
He extended the physical symmetry group to include gauge transformations, and showed that distinct spin liquids can be differentiated based on the behaviour of the gauge degrees of freedom.
We take this approach and apply it it to a continuum formulation of the phases in question.
However, as discussed, the true hallmark of a spin liquid is topological order, not the absence of broken symmetries, and there is no {\it a priori} reason to restrict to fully symmetric spin liquids.
We therefore also consider phases in which certain discrete symmetries are broken.

The physical symmetries relevant to the problem are the SU(2) spin symmetry, time reversal $\T$, and the space group symmetries.
The space group of the lattice is generated by the two translation operators, $T_x$ and $T_y$, the inversion operator $P_y$, and the rotation operator $R_{\pi/2}$. 
These act on the lattice sites as
\eq{\label{eqn:spaceGroup}
T_x&: (i_x,i_y)\mapsto (i_x+1,i_y),
&
T_y&: (i_x,i_y)\mapsto (i_x,i_y+1),
\nt 
P_y&: (i_x,i_y)\mapsto (i_x,-i_y),
&
R_{\pi/2}&: (i_x,i_y)\mapsto (-i_y,i_x).
}
In addition, these generators imply a symmetry under inversion of the $x$-coordinate, $P_x=R_{\pi/2}P_yR_{\pi/2}^{-1}$, as well as reflection about the $x=y$ axis, $P_{xy}=P_yR_{\pi/2}^{-1}$.
An equivalent definition of the space group is given through its commutation relations:
\eq{\label{eqn:spaceGpRels}
T_y^{-1}T_xT_yT_x^{-1}&=\id,
&
P_y^{-1}R_{\pi/2}P_yR_{\pi/2}&=\id,
\nt 
P_y^{-1}T_xP_yT_x^{-1}&=\id,
&
R_{\pi/2}^4&=\id,
\nt 
P_y^{-1}T_yP_yT_y&=\id,
&
R_{\pi/2}^{-1}T_xR_{\pi/2}T_y&=\id,
\nt 
P_y^2&=\id,
&
R_{\pi/2}^{-1}T_yR_{\pi/2}T_x^{-1}&=\id.
}
The generators all commute with time reversal, $G^{-1}\T^{-1}G\T=\id$, $G=\{T_x,T_y,P_y,R_{\pi/2}\}$. 
Because the fundamental degrees of freedom are bosonic spins, we have $\T^2=\id$.

Naturally, a different set of commutation relations is required to describe the space group in a symmetry broken phase, and these will be presented as needed.
To make contact with these phases, we will often describe the action of $P_x$ independently from the other symmetries even when considering fully symmetric spin liquids.

\subsection{Heisenberg antiferromagnet and the $\pi$-flux state}
\label{sec:PiFluxAnsatz}
We now present a lattice derivation of the $\pi$-flux model.
We begin by re-writing the spin operators in terms of so-called slave fermions \cite{Wen02}:
\eq{\label{eqn:slaveFermDef}
\v{S}_\vi&={1\o2}f_{\vi\a}^\dag\v{\s}_{\a\b}f_{\vi\b},
}
where $\v{\s}=\(\s^x,\s^y,\s^z\)$ are the Pauli matrices.
This expression introduces additional degrees of freedom and therefore cannot reproduce the Hilbert space of the spin operators without being supplemented by a constraint.
It can easily be verified that provided $\sum_\a f_{\vi\a}^\dag f_{\vi\a} =1$ on \emph{every} site, the representation in Eq.~\eqref{eqn:slaveFermDef} is correct. 
This further implies that $\sum_{\a,\b}\ep_{\a\b}f_{\vi\a}f_{\vi\b}=\sum_{\a,\b}\ep_{\a\b}f_{\vi\a}^\dag f_{\vi\b}^\dag=0$ where $\ep_{\a\b}$ is the fully anti-symmetric 2-index tensor.
By defining a matrix
\eq{
\mathcal{X}_\vi&=
\begin{pmatrix}
f_{\vi\uparrow} &   -f^\dag_{\vi\da} \\
f_{\vi\da}  &   f^\dag_{\vi\ua} 
\end{pmatrix}
}
we see that these constraints generate an SU(2) gauge symmetry which acts on $\mathcal{X}_\vi$ as
\eq{\label{eqn:LatticeGaugeAction}
\mathrm{SU}(2)_g : \mathcal{X}_\vi \to \mathcal{X}_\vi U_{g,\vi}^\dag.
}
The physical spin symmetry acts on $\mathcal{X}_\vi$ on the left:
\eq{
\mathrm{SU}(2)_s:   \mathcal{X}_\vi \to U_{s}\mathcal{X}_\vi.
}
The absence of a charge degree of freedom suggests that a more natural fermionic representation may be obtained by replacing the complex $f$-fermions with Majoranas:
\eq{
f_{\vi\uparrow}&={1\o\sqrt{2}}\(\chi_{\vi,0}+i\chi_{\vi,z}\),
&
f_{\vi\downarrow}&={1\o\sqrt{2}}\(-\chi_{\vi,y}+i\chi_{\vi,x}\),
}
where $\chi_{\vi,a}^\dag=\chi_{\vi,a}$ and $\{\chi_{\vi,a},\chi_{\vj,b}\}=\d_{ab}\d_{\vi\vj}$.
In this notation, the matrix $\mathcal{X}_\vi$ is written $\mathcal{X}_\vi={1\o\sqrt{2}}\(\chi_{\vi,0}+i\chi_{\vi,a}\s^a\)$ and the local constraints can be expressed as the conditions
\eq{\label{eqn:SU2GaugeConstraint}
\tr\(\s^a \mathcal{X}_\vi^\dag \mathcal{X}_\vi\)=0.
}


The first step to an approximate solution to $H_\mathrm{H}$ is to loosen the local constraint on the fermions to
\eq{\label{eqn:aveSU2GaugeConstraint}
\Braket{\tr\(\s^a \mathcal{X}_\vi^\dag \mathcal{X}_\vi\)}=0.
}
Next, we decouple the 4-fermion interaction through a Hubbard-Stratonovich transformation, leaving a quadratic mean field Hamiltonian.
The most general such Hamiltonian which can be made symmetric under spin rotation symmetry is \cite{Wen02,Chen2012}
\eq{\label{eqn:MFHamiltonian}
    H_\mathrm{MF}&=\sum_{\Braket{\vi\vj}}\bigg[
i \a_{\vi\vj}\tr\(\cX_\vi^\dag \cX_\vj\)
+ \b^a_{\vi\vj}\tr\(\s^a \cX_\vi^\dag \cX_\vj\)
+ i \g_{\vi\vj}\tr\(\s^a \cX_\vi^\dag \s^a\cX_\vj \)
\bigg],
}
where $\a_{\vi\vj}$, $\b^a_{\vi\vj}$, and $\g_{\vi\vj}$ are real numbers.
In accordance with its name, the $\pi$-flux  state is obtained by threading a $\pi$-flux through every plaquette: we take $\b^a_{\vi\vj}=\g_{\vi\vj}=0$ and
\eq{\label{eqn:SU2Ansatz}
\a_{\vi\vj}&=-\a_{\vj\vi},
&
\a_{\vi+\hx,\vi}&=\a,
&
\a_{\vi+\hy,\vi}&=(-1)^{i_x}\a.
}
This gives
\eq{\label{eqn:PiFluxHam}
H_{\pi}&=-i\a\sum_{\vi} 
\[
\tr\(\cX_{\vi}\cX_{\vi+\hx}\)+(-)^{i_x}\tr\(\cX_{\vi}\cX_{\vi+\hy}\)\].
}
While it is clear that this ansatz preserves the full SU(2) gauge and spin symmetries, the invariance of the $\pi$-flux Hamiltonian under the space group symmetries may be less clear.
In particular, translations in the $x$-direction do not preserve the form of $H_\pi$. 
However, the original Hamiltonian can be recovered through a gauge transformation, implying that the symmetry transformed state is (gauge) equivalent to the original.
Wen \cite{Wen02} termed this extended symmetry group the ``projective symmetry group" (PSG) and used it to show the existence of eight distinct fully-symmetric SU(2) spin liquids on the square lattice.
In his scheme, the Hamiltonian $H_\pi$ describes the SU2B$n0$ state (this is shown in Appendix~\ref{app:SU2Bn0latHam}). 
We will discuss the PSG extensively in subsequent sections, albeit in a slightly different context than originally formulated. 
His scheme is briefly reviewed in Appendix~\ref{app:WenClassification}.

The band structure of $H_\pi$ has two Dirac cones. 
We expand about these cones, labelling them by a valley index $v=1,2$.
A convenient expression for the resulting theory is achieved by defining the $4\times2$ matrix operator
\eq{
X_{\a,v;\b}&={1\o\sqrt{2}}\(\chi_{0,v}\d_{\a\b}+i\chi_{a,v}\s^a_{\a\b}\),
}
where $\a$, $\b$, and $v$ are spin, gauge, and valley indices respectively.
The low energy excitation of $H_\mathrm{MF}$ are described by the relativistic Dirac Lagrangian
\eq{
\L_\mathrm{MF}&=i\tr\( \bX \g^\m \ptl_\m X\)
}
where $\bch=\chi^T\g^0$, $\(\g^0,\g^x,\g^y\)=\(\t^y,i\t^z,i\t^x\)$. 
Here and in what follows, we express operators in \emph{real} time.

While Eq.~\eqref{eqn:aveSU2GaugeConstraint} may hold in the ground state of $H_{\pi}$, the full constraint in Eq.~\eqref{eqn:SU2GaugeConstraint} does not, and gauge fluctuations must be included to take this into account.
The SU(2) gauge transformation in Eq.~\eqref{eqn:LatticeGaugeAction} becomes
\eq{
\mathrm{SU(2)}_g:\;
X\rightarrow XU_g^\dag,\quad
a_\m\rightarrow U_ga_\m U_g^\dag+i\ptl_\m U_g U_g^\dag,
}
in the continuum.
As before, global spin rotations act the Majorana $X$ on the left,
\eq{
\mathrm{SU(2)}_s:X\rightarrow U_sX.
}
Letting $D_\m^aX=\ptl_\m X+iXa_\m,$ the inclusion of quantum fluctuations results in the following Lagrangian:
\eq{\label{eqn:QCDLagrangian}
\L_\qcd&=i\tr\(\bar{X}\g^\m D_\m^aX\).
}
$\L_\qcd$ can be expressed in a more familiar form by defining Dirac fermions
\eq{\label{eqn:DiracMajRelation}
\psi_{1,v}&={i\o\sqrt{2}}\(\chi_{x,v}-i\chi_{y,v}\),
&
\psi_{2,v}&=-{1\o\sqrt{2}}\(\chi_{0,v}+i\chi_{z,v}\).
}
In terms of these operators, the Lagrangian becomes
\eq{\label{eqn:LqcdDirac}
\L_\qcd&=\sum_{v=1,2}i\bpsi_v\g^\m\(\ptl_\m-ia_\m^a\s^a\)\psi_v.
}
That is, the low energy physics of the $\pi$-flux state is described by QCD$_3$ with $N_f=2$ fermions. 
The Dirac representation is not nearly as useful as the Majorana representation of $\L_\qcd$: while gauge transformations act of the $\psi$-fermions in the usual fashion,  the action of the spin symmetry is nontrivial.
We will therefore primarily use the form given in Eq.~\eqref{eqn:QCDLagrangian}.

A side-effect of the expansion about the Dirac cones is that the $\chi$ fermions transform nontrivially under time reversal and the space group symmetries:
\eq{\label{eqn:SymmTrans}
T_x&:\chi\rightarrow\m^x\chi,
&
R_{\pi/2}&:\chi\rightarrow e^{i\pi\t^y/4}e^{-i\pi\m^y/4}\chi(-y,x),
\nt
T_y&:\chi\rightarrow\m^z\chi,
&
P_x&:\chi\to \t^z\m^z \chi(-x,y),
\nt
\mathcal{T}&:\chi\rightarrow\t^y\m^y\chi,\quad i\rightarrow-i,
&
P_{y}&:\chi\to-\t^x\m^x\chi(x,-y).
}
In addition, the spin and space group symmetries of the model are significantly enlarged at this fixed point.
Not only is $\L_\qcd$ Lorentz invariant, but it is symmetric under rotations mixing the spin and valley indices of $X$: $X\rightarrow LX$, where $L$ is a $4\times4$ unitary matrix.
Because $X$ is composed of Majorana fermions, there is an important reality condition,
\begin{equation}\label{eqn:RealityCondition}
    X^*=\s^yX\s^y\,,
\end{equation} 
and therefore only $L$ such that $L^T\s^y L=\s^y$ are allowed. 
This reduces what would have been a U(4) symmetry to Sp(4). 
Finally, since both SU(2)$_g$ and Sp(4) share the nontrivial element $-1$, the true global symmetry is obtained by taking the quotient: $\mathrm{Sp}(4)/\Zt\cong\mathrm{SO}(5)$. 

\subsection{Dual description}\label{sec:duality}

As with any mean field approach involving a continuous gauge group, the existence of $\L_\qcd$ is by no means guaranteed once gauge fluctuations have been taken into account. 
However, in spite of some of the terminology, in this paper we do not view the $\pi$-flux `phase' as a stable state of matter existing over a finite region in parameter space.
Instead we treat it as a parent theory with instabilities potentially leading to U(1) and $\Zt$ spin liquids, as well as to ordered phases like N\'{e}el and VBS.
This approach is motivated by a duality between $\L_\qcd$ and $\mathds{CP}^1$ proposed by Wang {\it et al.} \cite{Wang17} to describe the N\'{e}el-VBS transition.
We discuss the relation between $\mathds{CP}^1$ and QCD$_3$ in this context.

One of the key components to their proposal is the SO(5) symmetry we just discussed.
On the QCD$_3$ side of the duality, an order parameter for this symmetry is
\eq{\label{eqn:SO5op}
n^j&=\tr\(\bX \Gamma^j X\),
&
\Gamma^j&=\{ \m^x,\m^z,\m^y\s^x,\m^y\s^y,\m^y\s^z\}.
}
The symmetry transformations in Eq.~\eqref{eqn:SymmTrans} indicate that $n^1$ and $n^2$ are the VBS order parameters, while $n^3$, $n^4$, and $n^5$ correspond to the N\'{e}el order parameter.
Using this, Refs.~\onlinecite{Tanaka05,Senthil06,Abanov00} showed that taking $\L_\qcd$ to
\eq{
\L_{\qcd,\phi}&=\L_\qcd + m\, \phi^j \,\tr\(\bX \Gamma^j X\),
}
and subsequently integrating out the fermions, yields a non-linear sigma model for $\phi$ with a Wess-Zumino-Witten (WZW) term.
This topological term manifests itself physically by making the defects of the order parameter of one symmetry transform nontrivially under the action of the other symmetry. 
These nontrivial correlations prompted Tanaka and Hu \cite{Tanaka05} and Senthil and Fisher \cite{Senthil06} to propose this non-linear sigma model as a description of the critical theory describing the Landau-forbidden continuous phase transition between N\'{e}el and VBS.

Conversely, the $\mathds{CP}^1$ formulation of the phase transition circumvents the obstruction to continuity by eschewing the traditional notion of an order parameter.
While the N\'{e}el phase is entered through the condensation of $N^a=z^\dag\s^a z$, the VBS phase is described by the proliferation of monopoles, events which change the flux of the gauge field by $2\pi$ (or, equivalently, change the global skyrmion number by one).
Not only do these monopoles confine the U(1) gauge field, but, because they transform nontrivially under the space group, this symmetry is necessarily broken in the condensate. 
In spite of the very different forms the N\'{e}el and VBS order parameters take, numerics \cite{Nahum15} have observed an emergent SO(5) symmetry between the two, implying that SO(5) emerges as a symmetry in the IR.
In this version, the VBS portion of SO(5) order parameter is given by $\(\phi_1,\phi_2\)=2\(\Re{\M},\Im{\M}\)$ where $\M$ denotes the monopole operator, while the remaining pieces are simply $\(\phi_3,\phi_4,\phi_5\)=\(z^\dag \s^x z,z^\dag \s^y z,z^\dag \s^z z\)$.

Wang et. al. \cite{Wang17} suggest that both of these models flow to the same SO(5) symmetric CFT in the IR. 
An important feature of this CFT is the absence of a relevant singlet operator.
The critical point is instead obtained by tuning the coupling $\m$ of a relevant, anistropic operator to zero, 
\eq{\label{eqn:AnisotropicOp}
\L&=\L_{\mathrm{SO}(5)}+\m\O_{an}
&
\O_{an}&\sim {2\o5}\(\phi_3^2+\phi_4^2+\phi_5^2\)-{3\o5}\(\phi_1^2+\phi_2^2 \).
}
When $\m>0$, the system has VBS order, while when $\m<0$, it orders along the N\'{e}el directions.
The approach we take is slightly different in spirit to this proposal, and we discuss this further in Sec.~\ref{sec:BerryPhase}.

\section{Spin liquids proximate to the $\pi$-flux
phase}\label{sec:ProximateSpinLiquids}
In this section, we describe the Higgs descendants of $\qcd$ and our approach to their classification.
We start by discussing which operators can couple to the Higgs field, before turning to a more complete discussion of the projective symmetry group than what was provided in the previous section.
Given a set of criteria described below, we conclude that there exists a single (spin) gapped U(1) spin liquid among the Higgs descendents of $\qcd$. 
We next list all gapped and fully symmetry $\Zt$ spin liquids, as well as all gapped $\Zt$ spin liquids with Ising-nematic order.
Special note is taken of the spin liquids $A_f$ and $B_f$, though we wait until until Sec.~\ref{sec:PSGcorres} to prove their equivalence to $A_b$ and $A_f$.
The section finishes with a description of the gapped $\Zt$ spin liquid with current-loop order we call $C_f$.

\subsection{Higgs fields}
\label{sec:Higgsf}

We being by examining the set of operators we will be coupling to the Higgs field. 
QCD$_3$ is strongly coupled in the IR, and so very little can be said with certainty regarding the operators and their scaling dimensions in the IR. 
We focus on fermion bilinears since these are the most relevant gauge invariant bosonic operators of the UV theory.
Non-perturbative operators such as monopoles are not considered.

We consider interaction terms of the form 
\eq{
\tr\(\vphi\bX M X\)= \vphi^a\tr\(\s^a\bX M X\)
}
where $\vphi=\vphi^a\s^a$ is a generic Higgs field transforming in the adjoint representation of SU(2)$_g$ and $M$ is a matrix acting on the sublattice, colour, and/or flavour space of the fermions and which may or may not contain derivatives.  
The physical properties of the various possible Higgs phases are defined primarily by the bilinear it couples to.

Restricting for the moment to bilinears without derivatives, those which are charged under the gauge group are
\eq{\label{eqn:GVecBilinNoDer}
&\tr\(\s^a\bX \g^\m X\),
&
&\tr\(\s^a\bX \Gamma^j \g^\m X\),
&
&\tr\(\s^a \bX T^j X \).
}
where $\Gamma^j=\{\m^z,-\m^x,\m^y\s^a\}$ and $T^j=\{\m^y,\s^a,\m^x\s^a,\m^z\s^a\}$ are the vector and adjoint representations of SO(5) respectively.
The first set of operators are the gauge currents $J^{a,\m}$. 
These cannot couple a Higgs field since the gauge theory description of the Heisenberg model is predicated on the requirement that these currents vanish.
In fact, the gauge fields can be interpreted as Lagrange multipliers which have been added to $\L_{\mathrm{QCD}_3}$ in order to impose the $J^{a,\m}=0$ constraint.

No such obstacles exist for the other two sets of bilinears.
The second group of operators,
$\tr\(\s^a\bX \Gamma^a \g^\m X\)$, are SO(5) and spacetime vectors in addition to gauge adjoints.
The presence of the gamma matrices $\g^\m$ indicates that the fermions will remain massless upon coupling these bilinears to a condensed $\vphi$.

On the other hand, should the Higgs field couple to one of the final operators in Eq.~\eqref{eqn:GVecBilinNoDer}, $\Braket{\vphi}\neq0$ will act as a mass for the fermions.
The only other bilinears which act as masses to the fermions are the singlet and SO(5) vector, neither of which are fully symmetric.
Therefore, given the aforementioned restriction on which operators we consider, we conclude that an operator of the form $\tr\(\s^a\bX T^j X\)$ must couple to a condensed Higgs field
in $A_f$ and $D_f$. 
(It can also be verified that these colour-singlet mass terms cannot provide a spin gap to the ordered spin liquids, $B_f$ or $C_f$.)


We will see shortly that the operators in Eq.~\eqref{eqn:GVecBilinNoDer} are not sufficient to reproduce the phase diagram in Figs.~\ref{fig:1}b and~\ref{fig:2}b.
Consequently, we also allow the Higgs field to couple to bilinears which contain a single derivative:
\eq{\label{eqn:bilinDer}
&\tr\(\s^a \bX i\ptl_\m X\),
&
&\tr\(\s^a \bX \Gamma^ji\ptl_\m X\),
&
&\tr\(\s^a \bX T^j\g^\m i\ptl_\n X\).
}

We now discuss how symmetries manifest in Higgs phases.
The action of the space group and time reversal on the bilinears listed above is given in Tables~\ref{tab:symmGenNoDer} and~\ref{tab:symmsVecNoDer}; the spin symmetry rotates operators with spin indices among themselves in the usual way.
It na\"{i}vely appears that a Higgs field coupling to any of these bilinears will necessarily break one or more symmetries upon condensing.
As with the $\pi$-flux Hamiltonian in Eq.~\eqref{eqn:PiFluxHam}, $H_\pi$, this intuition does not account for the fact that the Higgs field is not a gauge invariant operator. 
A symmetry is only truly broken if the original and symmetry transformed actions are \emph{not} gauge equivalent.

For instance, in Eq.~\eqref{eqn:HiggsLagrangian1}, $\tr\(\s^a\bX \m^y X\)$ couples to the Higgs field $\Phi$.
Since $\tr\(\s^a\bX \m^yX\)$ maps to minus itself under $\T$, $T_x$, and $T_y$, the na\"{i}ve argument would suggest that these symmetries are broken when $\Braket{\Phi^a}\neq0$.
However, it's not difficult to find a gauge transformation capable of ``undoing" the action of these symmetries.
In particular, supposing that only $\Braket{\Phi^x}\neq0$, we see that the gauge transformation $V=i\s^z$ takes $\tr\(\s^x \bX \m^y X\)$ to minus itself, thereby proving the equivalence of the original and symmetry transformed actions.

This set of gauge transformations comprises the PSG and is what we use to characterize the Higgs descendants.
More generally, when a group element acts on a bilinear as
\eq{
G: \tr\(\s^a \bX M X\) \to \tr\(\s^a \bX \bar{U}_G M U_G X\) 
}
where $\bar{U}_G=\g^0U^\dag\g^0$, the projective symmetry group is defined as
\eq{
\mathscr{P}G:\tr\(\s^a \bX M X\) \to  \tr\(V^\dag_G \s^a V_G \bX \bar{U}_G M U_G X\)
}
where
\eq{\label{eqn:PSGconstraint}
\tr\(V^\dag_G \s^a V_G \bX \bar{U}_G M U_G X\)=\tr\(\s^a \bX M X\).
}
We will see that requiring the existence of a $V_G$ for every $U_G$ places stringent conditions on which operators can couple to a Higgs field while preserving certain symmetries in the condensed phase.


\begin{table}
\centering
{\footnotesize
\begin{tabular}{|r|| c | c | c | c | c | r |}
\hlinewd{.8pt}
\multicolumn{1}{|c||}{$T^j$} & $\mathcal{T}$ & $P_{x}$ & $P_{y}$ & $T_x$ & $T_y$ & \multicolumn{1}{c|}{$R_{\pi/2}$}  \\\hlinewd{.8pt}\hline
\rowcolor{SeaGreen!15}
 $\m^y$ & $-$ & $+$ & $+$ & $-$ & $-$ & $\m^y$
  \\\hline
 $\s^a$ & $-$ & $-$ & $-$ & $+$ & $+$ & $\s^a$ 
  \\\hline
 $\m^x\s^a$  & $+$ & $+$ & $-$ & $+$ & $-$ & $\m^z\s^a$ 
  \\\hline
 $\m^z\s^a$ & $+$ & $-$ & $+$ & $-$ & $+$ & $-\m^x\s^a$
\\\hlinewd{.8pt}
\end{tabular}}
\caption{How $\tr\(\s^a\bX T^j X\)$ transform under the physical symmetries. $T^j=\{\m^y,\s^a,\m^x\s^a,\m^z\s^a\}$ are the 10 generators of SO(5).}
\label{tab:symmGenNoDer}
\end{table}
\begin{table}
\centering
{\footnotesize
\begin{tabular}{|r|| c | c | c | c | c | r |}
\hlinewd{.8pt}
\multicolumn{1}{|c||}{$\Gamma^j$} & $\mathcal{T}$ & $P_{x}$ & $P_{y}$ & $T_x$ & $T_y$ & \multicolumn{1}{c|}{$R_{\pi/2}$}
\\\hlinewd{.8pt}\hline
 $\m^x\g^0$ & $+$ & $-$ & $+$ & $+$ & $-$ & $\m^z\g^0$
  \\\hline
 $\m^x\g^x$  & $-$ & $+$ & $+$ & $+$ & $-$ & $\m^z\g^y$ 
  \\\hline
 $\m^x\g^y$ & $-$ & $-$ & $-$ & $+$ & $-$ & $-\m^z\g^x$ 
  \\\hlinewd{.8pt}
   $\m^z\g^0$ & $+$ & $+$ & $-$ & $-$ & $+$ & $-\m^x\g^0$
  \\\hline
 $\m^z\g^x$  & $-$ & $-$ & $-$ & $-$ & $+$ & $-\m^x\g^y$
  \\\hline
 $\m^z\g^y$ & $-$ & $+$ & $+$ & $-$ & $+$ & $\m^x\g^x$
  \\\hlinewd{.8pt}
 $\m^y\s^a\g^0$ & $+$ & $-$ & $-$ & $-$ & $-$ & $\m^y\s^a\g^0$
  \\\hline
 $\m^y\s^a\g^x$  & $-$ & $+$ & $-$ & $-$ & $-$ & $\m^y\s^a\g^y$ 
  \\\hline
 $\m^y\s^a\g^y$ & $-$ & $-$ & $+$ & $-$ & $-$ & $-\m^y\s^a\g^x$ 
  \\\hlinewd{.8pt}
\end{tabular}}
\caption{How $\tr\(\s^a\bX \Gamma^j \g^\m X\)$ transform under the physical symmetries. $\Gamma^j=\{\m^x,\m^z,\m^y\s^a\}$ transform under the vector representation of the emergent SO(5).}
\label{tab:symmsVecNoDer}
\end{table}
\definecolor{lightgray}{HTML}{EFEFEF}
\begin{table}
\centering
{\footnotesize
\begin{tabular}{|c|| c | c | c | c | c | c ||}
\hlinewd{.8pt}
$\m^y\g^\m i\ptl_\n$  & $\mathcal{T}$ & $P_{x}$ & $P_{y}$ & $T_x$ & $T_y$ & $R_{\pi/2}$ 
\\
 \hlinewd{.8pt}\hline
  \rowcolor{SkyBlue!15}
   $\m^y\g^0i\ptl_0$ & $+$ & $-$ & $-$ & $-$ & $-$ & $\ph{-}\m^y\g^0\ptl_0$ 
  \\\hline
 $\m^y\g^0i\ptl_x$ & $-$ & $+$ & $-$ & $-$ & $-$ & $\ph{-}\m^y\g^0\ptl_y$ 
  \\\hline
 $\m^y\g^0i\ptl_y$ & $-$ & $-$ & $+$ & $-$ & $-$ & $-\g^0\m^y\ptl_x$ 
   \\\hlinewd{.8pt}
 $\m^y\g^xi\ptl_0$ & $-$ & $+$ & $-$ & $-$ & $-$ & $\ph{-}\m^y\g^y\m^\ptl_0$ 
  \\\hline
   \rowcolor{RoyalBlue!15}
  $\m^y\g^xi\ptl_x$ & $+$ & $-$ & $-$ & $-$ & $-$ & $\ph{-}\m^y\g^yi\ptl_y$ 
  \\\hline
  \rowcolor{magenta!15}
 $\m^y\g^xi\ptl_y$ & $+$ & $+$ & $+$ & $-$ & $-$ & $-\m^y\g^yi\ptl_x$ 
  \\\hlinewd{.8pt}
$\m^y\g^yi\ptl_0$ & $-$ & $-$ & $+$ & $-$ & $-$ & $-\m^y\g^xi\ptl_0$  
  \\\hline
   \rowcolor{magenta!15}
$\m^y\g^yi\ptl_x$ & $+$ & $+$ & $+$ & $-$ & $-$ & $-\m^y\g^xi\ptl_y$  
  \\\hline
    \rowcolor{RoyalBlue!15}
 $\m^y\g^yi\ptl_y$& $+$ & $-$ & $-$ & $-$ & $-$ & $\ph{-}\m^y\g^xi\ptl_x$ 
   \\\hline
\end{tabular}
\hspace{3mm}
\begin{tabular}{|c|| c | c | c | c | c | c |}
\hlinewd{.8pt}
$\m^{0,x,y}i\ptl_\m$  & $\mathcal{T}$ & $P_{x}$ & $P_{y}$ & $T_x$ & $T_y$ & $R_{\pi/2}$ \\\hlinewd{.8pt}\hline
  \rowcolor{violet!15}
    $i\ptl_0$& $+$ & $-$ & $-$ & $+$ & $+$ & $\ph{-}i\ptl_0$
  \\\hline
  $i\ptl_x$& $-$ & $+$ & $-$ & $+$ & $+$ & $\ph{-}i\ptl_y$
  \\\hline
  $i\ptl_y$& $-$ & $-$ & $+$ & $+$ & $+$ & $-i\ptl_x$
   \\\hlinewd{.8pt}
 $\m^xi\ptl_0$ & $-$ & $+$ & $-$ & $+$ & $-$ & $\ph{-}\m^zi\ptl_0$ 
  \\\hline
  $\m^xi\ptl_x$ & $+$ & $-$ & $-$ & $+$ & $-$ & $\ph{-}\m^zi\ptl_y$
  \\\hline
 $\m^xi\ptl_y$ & $+$ & $+$ & $+$ & $+$ & $-$ & $-\m^zi\ptl_x$ 
  \\\hlinewd{.8pt}
$\m^zi\ptl_0$ & $-$ & $-$ & $+$ & $-$ & $+$ & $-\m^xi\ptl_0$ 
  \\\hline
$\m^zi\ptl_x$ & $+$ & $+$ & $+$ & $-$ & $+$ & $-\m^xi\ptl_y$ 
  \\\hline
 $\m^zi\ptl_y$ & $+$ & $-$ & $-$ & $-$ & $+$ & $\ph{-}\m^xi\ptl_x$ 
   \\\hlinewd{.8pt}
\end{tabular}
}
\caption{Symmetry transformation properties of bilinears of the form $\tr\(\s^a\bX i \ptl_\m X\)$, $\tr\(\s^a\bX \Gamma^ji\ptl_\m X\)$, and $\tr\(\s^a\bX T^j\gamma^\m i\ptl_\n X\)$ which do \emph{not} transform under spin.
The operators which can couple to a Higgs fields in a gapped symmetric spin $\Zt$ spin liquid are coloured; entries with the same colour transform into one another under $R_{\pi/2}$.}
\label{tab:symmsNonVan2}
\end{table}
\subsection{Symmetric spin liquids}
\label{sec:symmliq}

In this section, we focus on fully symmetric and gapped spin liquids (by `gapped,' we are referring specifically to the matter content).
As mentioned, in order to simultaneously gap the fermions and Higgs the gauge boson, an operator of the form $\tr\(\s^a\bX T^j X\)$ where $T^j$ is a generator of SO(5) must couple to a Higgs field.
These are listed in Table~\ref{tab:symmGenNoDer}.
Of the ten generators of SO(5), nine transform as vectors under the spin symmetry, and we show in Appendix~\ref{app:ProjSpinSymm} that a fully symmetry spin liquid cannot be formed by coupling a Higgs field to any of these bilinears. 
Roughly, the argument relies on the fact that in order to preserve the spin symmetry, a linear combination of the form $\sim\sum_a \tr\( \s^a \bX M \s^a X\)$ for $M=\id,\m^x,\m^z$ must couple to the Higgs, which then makes it impossible preserve all of the discrete symmetries.

This observation establishes $\tr\(\s^a \bX \m^y X \)$ as the only fermion bilinear capable of both giving the fermions a mass and coupling to a Higgs field.
As indicated in Eq.~(\ref{f2}) and Section~\ref{sec:Higgsf}, 
we denote the Higgs field coupling to this bilinear as $\Phi^a$.
Since the action remains invariant under all gauge transformations about the direction of the condensate, $\Phi^a$ cannot fully Higgs the SU(2) gauge symmetry down to $\Zt$.
For instance, if we will assume that only $\Braket{\Phi^x}\neq0$, U(1) operations of the form $X\rightarrow X e^{-i\th \s^x}$ remain a gauge symmetry.
We label this U(1) spin liquid $D_f$.

It is well-known \cite{Hermele04} 
that without gapless degrees of freedom, a U(1) gauge theory is unstable to the proliferation of monopoles and confinement \cite{Polyakov75}. 
We will ignore the ultimate fate of $D_f$ until Sec.~\ref{sec:BerryPhase} where we show that the true ground state is a VBS.

With this caveat in mind, we deduce the projective symmetry group 
of the gapped U(1) spin liquid from Table~\ref{tab:symmGenNoDer}:
\eq{\label{eqn:U1GappedPSGCont}
V_t&=e^{i\th_t\s^x}i\s^z,
&
V_{tx}&=e^{i\th_{tx}\s^x}i\s^z,
\nt
V_{py}&=e^{i\th_{py}\s^x},
&
V_{ty}&=e^{i\th_{ty}\s^x}i\s^z,
\nt
V_{px}&=e^{i\th_{py}\s^x},
&
V_{r}&=e^{i\th_r\s^x},
}
where the $\th_G$ are arbitrary angles parametrizing the residual U(1) gauge degree of freedom.
Here, the subscripts $t,px,py,tx,ty$ and $r$ indicate that these gauge transformation accompany the action of $\T,P_x,P_y,T_x,T_y$, and $R_{\pi/2}$ respectively.

We note that while the physical symmetries are all preserved in $D_f$, the emergent SO(5) symmetry of $\qcd$ has been broken.
Of the SO(5) generators, $T^j=\{\m^y,\s^a,\m^x\s^a,\m^z\s^a\}$, the U(1) gauge theory is only invariant under $\{\m^y\}\times \{\s^a\}$, indicating that the SO(5) is broken to U(1)$\times$SU(2).
From the perspective of the SO(5) order parameter, $n^j=\tr\(\bX \Gamma^j X\)$, $\Gamma^j=\{\m^x,\m^z,\m^y\s^a\}$, the VBS order parameters, $n^1$ and $n^2$ can no longer be rotated into the N\'{e}el order parameters, $n^3$, $n^4$, and $n^5$.

To break the gauge group down to $\Zt$, an additional Higgs field $\Phi_1$ is needed. 
However, there are strict constraints on which bilinears can couple to $\Phi_1$ in order for the resultant $\Zt$ spin liquid to preserve all physical symmetries. 
We approach this problem from a vector representation by associating an SO(3) matrix $Q$ to each SU(2)$_g$ gauge transformation $V$.
That is, instead of looking at $V$ such that $\tr\(\vphi \bX M X\)\to \tr\( V^\dag\vphi V \bX M X\)$, we consider $Q$ such that
$\vphi^a\tr\(\s^a \bX M X\)\to (Q\vphi)^a\tr\( \s^a \bX M X\)$.
In this notation, when $\Braket{\Phi^x}\neq0$, we must have
\eq{
Q_G&=
\begin{pmatrix}
1 &   \v{0}  \\
\v{0}   & R_G 
\end{pmatrix},
&
G&=px,py,r,
}
and
\eq{
Q_G&=
\begin{pmatrix}
-1  &   \v{0}  \\
\v{0}   &   \tilde{R}_G 
\end{pmatrix},
&
G&=t,tx,ty,
}
where $R_G$ and $\tilde{R}_G$ are determined by the bilinear coupling to $\Phi_1$. 
The constraints on this bilinear arise from the fact that $Q_G$ must be special orthogonal, therefore implying that $R_G$ and $\tilde{R}_G$ must be $2\times2$ orthogonal matrices with determinants $+1$ and $-1$ respectively.

We now argue that none of the operators in Table~\ref{tab:symmsVecNoDer} satisfy these requirements.
First, all bilinears with spin indices can be excluded by the same reasoning given above and in Appendix~\ref{app:ProjSpinSymm}.
Next, we note that all remaining operators still transform differently than $\tr\(\s^a\bX \m^y X\)$ under at least one of the symmetries, and therefore the $\Phi_1$ condensate \emph{must} be perpendicular to $x$ in colour space.
For the remaining six operators, the obstruction to forming a spin liquid may be understood by studying the action of a 90$^0$ rotation. 
The last column of the table indicates that $R_{\pi/2}$ maps each bilinear to plus or minus another bilinear in the table, eg. $R_{\pi/2}: \tr\(\s^a \bX \m^x\g^0 X\)\to \tr\(\s^a \bX \m^z \g^0X\)$.
In order for this to describe a rotationally symmetric phase, both bilinears must couple to a Higgs field.
We might imagine that $\Phi_1$ couples to both operators in a pair,
but this is not a viable option because the other discrete symmetries do not act on the members of each pair in the same way. 
For instance, no gauge transformation can preserve the form of $\Braket{\Phi_1^a}\,\tr\(\s^a\bX \g^0\[\m^x\pm\m^z\] X\)$ 
under $P_x,P_y,T_x,$ and $T_y$ since
$\tr\(\s^a\bX \g^0 \m^x X\)$ and $\tr\(\s^a\bX \g^0\m^z X\)$ behave differently under these symmetries.
We might try coupling each of these operators to \emph{different} Higgs fields, $\Phi_1$ and $\Phi_1'$, and require that they condense in mutually perpendicular channels, eg. $\Braket{\Phi_1^y}\neq0$ and $\Braket{{\Phi_1'}^z}\neq0$.
However, the matrix required to undo the action of the time reversal symmetry is then $Q_t=\diag(-1,1,1)$ which is not an element of SO(3).
We conclude that this does not work either.

We next perform the same analysis on bilinears containing a single derivative.
Once again, the arguments in Appendix~\ref{app:ProjSpinSymm} are valid, and we immediately exclude all operators in Eq.~\eqref{eqn:bilinDer} which transform nontrivially under spin rotations.
The action of the space group and time reversal symmetries on the remaining operators is provided in Table~\ref{tab:symmsNonVan2}.
Again, $R_{\pi/2}$ maps many of the operators to plus or minus a different operator in the table. 
As discussed in the previous paragraph, only bilinears which transform in the same way under $\T,P_x,P_y,T_x$, and $T_y$ as their partner under $R_{\pi/2}$ are suitable candidates, and these have been highlighted in different colours.
In Table~\ref{tab:PSGs} we list the PSG's of all gapped and symmetric $\Zt$ spin liquids which can be formed using this set of operators.


In Sec.~\ref{sec:PSGcorres} we determine which (if any) bosonic ansatz these PSG's correspond to.
We find that $s$PSG5 corresponds to the fully symmetric spin liquid $A_b$, and for this reason, we denote it $A_f$.
\definecolor{lightgray}{HTML}{EFEFEF}
\begin{table}
\centering
{\footnotesize
\begin{tabular}{ | c || l || r |r |r |r |}
\hline
$s$PSG 	&	\multicolumn{1}{c||}{$M$}	&	\multicolumn{1}{c|}{$Q_t/V_t$}	&	\multicolumn{1}{c|}{$Q_{px,py}/V_{px,py}$}	&	\multicolumn{1}{c|}{$Q_{tx,ty}/V_{tx,ty}$}	&	\multicolumn{1}{c|}{$Q_r/V_r$}
\\\hlinewd{.8pt}\hline
	&		&	
	$\diag(-1,1,1)$	&	$\diag(1,-1,-1)$	&	$\diag(-1,-1,1)$	&	$\id$
	\\\cline{3-6}
\multirow{-2}{*}{1}	    & \multirow{-2}{*}{$\m^y\g^0i\ptl_0,\;\m^y\(\g^xi\ptl_x+\g^yi\ptl_y\)$}
    &   $i\s^y$ &   $i\s^x$ &   $i\s^z$ &   $\id$
\\\hlinewd{.8pt}
	&		&	
	$\diag(-1,1,1)$	&	$\diag(1,-1,-1)$	&	$\diag(-1,-1,1)$	&	$\diag(1,-1,-1)$
	\\\cline{3-6}
\multirow{-2}{*}{2} &   \multirow{-2}{*}{$\m^y\(\g^xi\ptl_x-\g^yi\ptl_y\)$}  &
    $i\s^y$ &   $i\s^x$ &   $i\s^z$ &   $i\s^x$
\\\hlinewd{.8pt}
	&		&	
	$\diag(-1,1,1)$	&	$\id$	&	$\diag(-1,-1,1)$	&	$\diag(1,-1,-1)$
    \\\cline{3-6}
\multirow{-2}{*}{3} &   \multirow{-2}{*}{$\m^y\(\g^xi\ptl_y+\g^yi\ptl_x\)$}  &
    $i\s^y$ &   $\id$   & $i\s^z$   &   $i\s^x$ 
\\\hlinewd{.8pt}
	&		&	
	$\diag(-1,1,1)$	&	$\id$	&	$\diag(-1,-1,1)$	&	$\id$
    \\\cline{3-6}
\multirow{-2}{*}{4} &   \multirow{-2}{*}{$\m^y\(\g^xi\ptl_y-\g^yi\ptl_x\)$}    &
    $i\s^y$ &   $\id$   &   $i\s^z$ &   $\id$   
\\\hlinewd{.8pt}
	\rowcolor{Yellow!40}
	&		&	
	$\diag(-1,1,1)$	&	$\diag(1,-1,-1)$	&	$\diag(-1,1,-1)$	&	$\id$
    \\\cline{3-6}
    \rowcolor{Yellow!40}
\multirow{-2}{*}{5} &   \multirow{-2}{*}{$i\ptl_0$} &  
    $i\s^y$ &   $i\s^x$ &   $i\s^y$ &   $\id$
\\\hlinewd{.8pt}
\end{tabular}}
\caption{All symmetric PSG's associated with symmetric $\Zt$ spin liquids in which $\Braket{\Phi^x}\neq0$ where $\Phi$ couples to $\tr\(\s^a \bX \m^y X\)$. These are listed as a function of the operator $\tr\(\s^a\bX M X\)$ which $\Phi_1$ couples to. We assume that only $\Braket{\Phi_1^y}\neq0$.
}
\label{tab:PSGs}
\end{table}

\subsection{$\Zt$ spin liquids with Ising-nematic order}
\label{sec:Isingnematic}

As emphasized in Section~\ref{sec:intro}, it is not necessary to restrict to fully symmetric spin liquids.
We therefore also study gapped, nematic $\Zt$ spin liquids proximate to the gapped U(1) spin liquid $D_f$.
In particular, we investigate spin liquids which are obtained by coupling a third Higgs field, $\Phi_{2i}$, to the operators in Tables~\ref{tab:symmGenNoDer}, \ref{tab:symmsVecNoDer}, and~\ref{tab:symmsNonVan2}, and which preserve the continuous spin symmetry, $\T$, $P_x$, $P_y$, $T_x$, and $T_y$, but break the 90$^0$ rotation symmetry, $R_{\pi/2}$.
The absence of rotation symmetry makes it possible to couple any of the operators in Tables~\ref{tab:symmGenNoDer} and~\ref{tab:symmsVecNoDer} to the Higgs field, and the ten candidates we find are listed in Table~\ref{tab:NematicPSGs}.

We note that $n$PSG5 and $n$PSG6 are continuously connected to $s$PSG1-2 and $s$PSG3-4 respectively. 
For instance, in the case of $n$PSG5, if the Higgs field couples as $\sum_{i=x,y}\Phi^a_{2i}\tr\(\s^z\bX \g^ii\ptl_i X\)$, then phases where the condensate satisfies $\Braket{\Phi^a_{2x}}=\pm\Braket{\Phi^a_{2y}}$ do not break $R_{\pi/2}$ and are precisely $s$PSG1 and $s$PSG2.
The same considerations hold for $n$PSG6 in relation to $s$PSG3 and $s$PSG4.

In all cases, the phase with $\Braket{\Phi}=0$ and $\Braket{\Phi_1}\neq0$ is a fully symmetric U(1) spin liquid. However, unlike $D_f$, the matter sector is gapless.

In the next section we find that $n$PSG7 is the fermionic version of the bosonic phase $B_b$, leading us to label it $B_f$. 

\begin{table}
\centering
{\small
\begin{tabular}{ | c | c || l || r | r | r | r | r |}
\hlinewd{.8pt}
\multicolumn{2}{|c||}{$n$PSG}	&	\multicolumn{1}{c||}{$M_i$}	&	\multicolumn{1}{c|}{$V_t$}	&	
\multicolumn{1}{c|}{$V_{px}$}	&	
\multicolumn{1}{c|}{$V_{py}$}	&	
\multicolumn{1}{c|}{$V_{tx}$}	&	
\multicolumn{1}{c|}{$V_{ty}$}	
\\\hlinewd{.8pt}\hline	
    &	\multirow{-1}{*}{$x$}   & 
    \multirow{-1}{*}{$\m^x\g^0$}  &
	$i\s^y$	&	$i\s^x$	&	$\id$	&	$i\s^y$	&	$i\s^z$
\\\cline{2-8}	
	\multirow{-2}{*}{1} & 
	\multirow{-1}{*}{$y$} &
	\multirow{-1}{*}{$\m^z\g^0$}    &
	$i\s^y$	&	$\id$		&	$i\s^x$	&	$i\s^z$	&	$i\s^y$
\\\hlinewd{.8pt}
	&   
	\multirow{-1}{*}{$x$} & 
	\multirow{-1}{*}{$\m^z\g^y$}    &
	$i\s^z$	&	$\id$		&	$\id$	&	$i\s^z$	&	$i\s^y$
	\\\cline{2-8}
	\multirow{-2}{*}{2}& 	
	\multirow{-1}{*}{$y$}   & 
	\multirow{-1}{*}{$\m^x\g^x$}	&
	$i\s^z$	&	$\id$	&	$\id$	&	$i\s^y$	&	$i\s^z$
\\\hlinewd{.8pt}
	& 
	\multirow{-1}{*}{$x$} & 
	\multirow{-1}{*}{$\m^x\g^y,$} 	&
	$i\s^z$	&	$i\s^x$	&$i\s^x$	&	$i\s^y$	&	$i\s^z$
\\\cline{2-8}	
	\multirow{-2}{*}{3} &
	\multirow{-1}{*}{$y$}   &
	\multirow{-1}{*}{$\m^z\g^x$}    &
	$i\s^z$	&	$i\s^x$		&	$i\s^x$	&	$i\s^z$	&	$i\s^y$
	\\\hlinewd{.8pt}
	&   
	\multirow{-1}{*}{$x$}  &
	\multirow{-1}{*}{$\m^y\g^0i\ptl_x,\,\m^y\g^xi\ptl_0$}	&
	$i\s^z$	&	$\id$		&	$i\s^x$	&	$i\s^z$	&	$i\s^z$
	\\\cline{2-8}	
	\multirow{-2}{*}{4}  &
	\multirow{-1}{*}{$y$}  &
	\multirow{-1}{*}{$\m^y\g^0i\ptl_y,\,\m^y\g^yi\ptl_0$}  &
	$i\s^z$	&	$i\s^x$		&	$\id$	&	$i\s^z$	&	$i\s^z$
\\\hlinewd{.8pt}
	&
	\multirow{-1}{*}{$x$}   &
	\multirow{-1}{*}{$\m^y\g^xi\ptl_x$}	&
	$i\s^y$	&	$i\s^x$		&	$i\s^x$	&	$i\s^z$	&	$i\s^z$
	\\\cline{2-8}		
	\multirow{-2}{*}{5} &	
	\multirow{-1}{*}{$y$}  &
    \multirow{-1}{*}{$\m^y\g^yi\ptl_y$}	&
	$i\s^y$	&	$i\s^x$		&	$i\s^x$	&	$i\s^z$	&	$i\s^z$
\\\hlinewd{.8pt}
	&
	\multirow{-1}{*}{$x$}   &
	\multirow{-1}{*}{$\m^y\g^xi\ptl_y$}	&
	$i\s^y$	&	$\id$		&	$\id$	&	$i\s^z$	&	$i\s^z$
	\\\cline{2-8}	
	\multirow{-2}{*}{6} &	
	\multirow{-1}{*}{$y$}  &
	\multirow{-1}{*}{$\m^y\g^yi\ptl_x$}	&
	$i\s^y$	&	$\id$		&	$\id$	&	$i\s^z$	&	$i\s^z$
\\\hlinewd{.8pt}
	\rowcolor{SkyBlue!20}
	&
	\multirow{-1}{*}{$x$}    &
	\multirow{-1}{*}{$i\ptl_x$}  &
	$i\s^z$	&	$\id$		&	$i\s^x$	&	$i\s^y$	&	$i\s^y$
	\\\cline{2-8}
	\rowcolor{SkyBlue!20}
	\multirow{-2}{*}{7} &	
	\multirow{-1}{*}{$y$}   &
	\multirow{-1}{*}{$i\ptl_y$}  &
	$i\s^z$	&	$i\s^x$		&	$\id$	&	$i\s^y$	&	$i\s^y$
\\\hlinewd{.8pt}
	&
	\multirow{-1}{*}{$x$}   &
	\multirow{-1}{*}{$\m^xi\ptl_0$}    &
	$i\s^z$	&	$\id$		&	$i\s^x$	&	$i\s^y$	&	$i\s^z$
	\\\cline{2-8}
	\multirow{-2}{*}{8} &		
	\multirow{-1}{*}{$y$}   &
	\multirow{-1}{*}{$\m^zi\ptl_0$}   &
	$i\s^z$	&	$i\s^x$		&	$\id$	&	$i\s^z$	&	$i\s^y$
	\\\hlinewd{.8pt}
	&
	\multirow{-1}{*}{$x$}   &
	\multirow{-1}{*}{$\m^zi\partial_x$}    &
	$i\s^y$	&	$\id$		&	$\id$	&	$i\s^z$	&	$i\s^y$
	\\\cline{2-8}
	\multirow{-2}{*}{9} &	
	\multirow{-1}{*}{$y$}   &
	\multirow{-1}{*}{$\m^xi\partial_y$}   &
	$i\s^y$	&	$\id$		&	$\id$	&	$i\s^y$	&	$i\s^z$
	\\\hlinewd{.8pt}
	&
	\multirow{-1}{*}{$x$}   &
	\multirow{-1}{*}{$\m^xi\partial_x$}    &
	$i\s^y$	&	$i\s^x$		&	$i\s^x$	&	$i\s^y$	&	$i\s^z$
	\\\cline{2-8}
	\multirow{-2}{*}{10} &
	\multirow{-1}{*}{$y$}   &
	\multirow{-1}{*}{$\m^zi\partial_y$}   &
	$i\s^y$	&	$i\s^x$		&	$i\s^x$	&	$i\s^z$	&	$i\s^y$
	\\\hlinewd{.8pt}
\end{tabular}}
\caption{Nematic PSG's associated with order parameters of the form $\Phi^a\tr\(\s^a\bX \m^y X\)+\Phi_{2i}^a\tr\(\s^a\bX M^i X\)$.
We have not included $\tr\(\s^a\bX \ptl_0 X \)$ since this operator is invariant under the action of $R_{\pi/2}$ and already accounted for as $s$PSG5. 
The labels $x$, $y$ are simply a convenient notation and do not necessarily signify a physical direction.}
\label{tab:NematicPSGs}
\end{table}
\subsection{$\Zt$ spin liquid with current-loop order}
\label{sec:currentloop}

So far, we have defined three separate Higgs fields.
To ensure that the condensed phases had a spin gap, $\Phi$ and $\tr\(\s^a\bX \m^y X\)$ were required to couple.
We then identified which bilinears could couple to a second Higgs field, $\Phi_1$, such that the phase with $\Braket{\Phi}\neq0$, $\Braket{\Phi_1}\neq0$, and $\Braket{\Phi}\perp\Braket{\Phi_1}$ was a fully symmetric spin $\Zt$ liquid.
Similarly, we determined in the previous section which bilinears could couple to a Higgs field $\Phi_{2i}$ such that the phase with $\Braket{\Phi}\neq0$ and $\Braket{\Phi_{2i}}\neq0$ was a $\Zt$ spin liquid with Ising-nematic order, again provided  $\Braket{\Phi}\perp\Braket{\Phi_{2i}}$.

A natural extension is to ask which phases result when all three Higgs fields have condensed: $\Braket{\Phi}\neq0$, $\Braket{\Phi_{1}}\neq0$, and $\Braket{\Phi_{2i}}\neq0$.
However, there are clearly a large number of possibilities. 
Not only have we identified many candidate $s$- and $n$PSG's, but different symmetries will be broken depending on the relative orientation of the Higgs fields. 
Therefore, we focus on producing the phase diagram in Fig.~\ref{fig:2}b and restrict our study to the situation where the symmetric and nematic spin liquids are $A_f$ and $B_f$, the phases described by $s$PSG5 and $n$PSG7.

This scenario describes four different patterns of symmetry breaking:
\begin{enumerate}
\item $\Braket{\Phi}\perp\Braket{\Phi_1},\Braket{\Phi}\perp\Braket{\Phi_{2i}}$, \& $\Braket{\Phi_1}\parallel \Braket{\Phi_{2i}}$
\item $\Braket{\Phi}\perp \Braket{\Phi_1},\Braket{\Phi_1}\perp\Braket{\Phi_{2i}}$, \& $\Braket{\Phi}\parallel\Braket{\Phi_{2i}}$
\item $\Braket{\Phi_{2i}}\perp \Braket{\Phi},\Braket{\Phi_{2i}}\perp\Braket{\Phi_{1}}$, \& $\Braket{\Phi}\parallel\Braket{\Phi_{1}}$
\item  $\Braket{\Phi}\perp\Braket{\Phi_1}\perp\Braket{\Phi_{2i}}$
\end{enumerate}
In Table~\ref{tab:3rdPhase} we list which symmetries are broken for each of these cases.

Referring to the phase diagram in Fig.~\ref{fig:1}b, it is natural to restrict to the case where $A_f$ and $B_f$ are accessible by taking $\Braket{\Phi_{2i}}$ or $\Braket{\Phi_1}$ to zero.
Since both the second and third cases have $\Braket{\Phi}$ parallel to either $\Braket{\Phi_1}$ or $\Braket{\Phi_{2i}}$, we eliminate these options.
Of the remaining two phases, the resulting spin liquid only possesses current-loop order when $\Braket{\Phi_1}\parallel\Braket{\Phi_{2i}}$.
This situation is further distinguished by breaking the fewest symmetries.
We refer to this phase as $C_f$ and later equate it and the bosonic phase $C_b$.

\begin{table}
\centering
\begin{tabular}{| c | c | c | c || c |}
\hlinewd{.8pt}
    &   \multicolumn{3}{c||}{Direction}  &   
\\\cline{2-4} 
    & $\Braket{\Phi}$   &   $\Braket{\Phi_1}$   &   $\Braket{\Phi_{2x}}$ &   broken    
\\\hlinewd{.8pt}\hline
\rowcolor{WildStrawberry!30}
1   &   $x$   &   $y$   &   $y$   &  
    $\T$, $P_x$ 
\\\hline
2   &   $x$   &   $y$   &   $x$   &  
    $P_y$, $T_x$, $T_y$   
\\\hline
3   &   $x$   &   $x$   &   $y$   &  
    $\T$, $P_x$, $P_y$, $T_x$, $T_y$     
\\\hline
4   &   $x$   &   $y$   &   $z$   &
    $P_x$, $T_x$, $T_y$   
\\\hlinewd{.8pt}
\end{tabular}
\caption{
Symmetries broken depending on the orientation in gauge space taken by the Higgs condensates. The fields couple to the bilinears as $\tr\(\Phi\bX \m^y X\)+\tr\(\Phi_1\bX i\ptl_0 X\)+ \tr\(\Phi_{2x}\bX i\ptl_xX \)$.}
\label{tab:3rdPhase}
\end{table}

\section{Spin liquid identification}\label{sec:SpinLiqID}
We now identify the phases examined above with previously studied spin liquids.
On the lattice, Wen \cite{Wen02} showed that 58 distinct $\Zt$ PSG's can be accessed from the $\pi$-flux state (SU2B$n0$).
However, his PSG classification gives no indication of the physical properties of these phases and, moreover, as we will see, it includes certain ``anomalous" PSG's which cannot be obtained from a mean field ansatz.
We therefore begin by discussing the ``symmetry fractionalization" approach to spin liquid classification, and relate it to Wen's scheme. 
This will significantly simplify the process of relating the symmetric U(1) spin liquids and the phases in Table~\ref{tab:PSGs} to the spin liquids studied by Wen.
Its greatest power, however, will be to treat fermionic and bosonic mean field ansatze on the same footing, allowing us relate our results to phases described using Schwinger bosons, and prove our earlier claim that $A_f,B_f$, and $C_f$ are fermionic versions of $A_b$, $B_b$, and $C_b$.

We next show that the gapped U(1) spin liquid $D_f$ corresponds to $D_b$.
The gapless gauge degrees of freedom invalidate the symmetry fractionalizaton approach to comparing spin liquids represented with bosons and fermions.
Instead, we show through linear response that the proliferation of monopoles induces the condensation of the VBS order parameters given by the first two components of the vector in Eq.~\eqref{eqn:SO5op}.
We provide additional verification by demonstrating that the Berry phase of the monopole matches the calculation performed by Haldane \cite{Haldane88} and Read and Sachdev \cite{NRSS90}.

\subsection{Symmetry fractionalization and $\Zt$ spin liquid identification}\label{sec:SymFrac}
\begin{table}
\centering
{\footnotesize
%
\begin{tabular}{|r| l|| r | r | r | r | >{\columncolor{Yellow!40}} r || r | r | r | r | r |}
\hline
& Group relations & $s$PSG1 &   $s$PSG2 &   $s$PSG3
&   $s$PSG4 &   $s$PSG5 & vison  & twist & $\Zt[0,0]$ & $\Zt[0,\pi]$
\\
\hline\hline
1	&	$T_y^{-1}T_xT_yT_x^{-1}$ 		&	$-1$	&	$-1$	&	$-1$	&	$-1$	&	$-1$	&	$-1$	& 	$1$	& 	$1$	&	$1$
\\	
2	&	$P_y^{-1}T_xP_yT_x^{-1}$		&	$-1$	&	$-1$	&	$1$	&	$1$	&	$-1$	&	$-1$	& 	$1$	&	$1$	&	$1$
\\
3	&	$P_y^{-1}T_yP_yT_y$			&	$-1$	&	$-1$	&	$1$	&	$1$	&	$-1$	&	1	&	$1$	&	$-1$	&	$-1$
\\
4	&	$P_y^2$						&	$-1$	&	$-1$	&	$1$	&	$1$	&	$-1$	&	1	&	$-1$	&	$1$	&	$1$
\\
5	&	$P_y^{-1}R_{\pi/2}P_yR_{\pi/2}$	&	$1$	&	$-1$	&	$-1$	&	$1$	&	$1$	&	1	&	$1$	&	$1$	&	$-1$
\\
6	&	$R_{\pi/2}^4$					&	$1$	&	$1$	&	$1$	&	$1$	&	$1$	&	$-1$	&	$-1$ &	$1$	&	$1$
\\
7	&	$R_{\pi/2}^{-1}T_xR_{\pi/2}T_y$	&	$-1$	&	$-1$	&	$-1$	&	$-1$	&	$-1$	&	$-1$	&	$1$	&	$1$	&	$1$
\\
8	&	$R_{\pi/2}^{-1}T_yR_{\pi/2}T_x^{-1}$	&	$-1$	&	$-1$	&	$-1$	&	$-1$	&	$-1$	&	1	&	$1$	&	$-1$	&	$-1$
\\
9	&	$R_{\pi/2}^{-1}\T^{-1}R_{\pi/2}\T$	&	$1$	&	$-1$	&	$-1$	&	$1$	&	$1$	&	1	&	$1$	&	$1$	&	$-1$
\\
10	&	$P_y^{-1}\T^{-1}P_y\T$			&	$-1$	&	$-1$	&	$1$	&	$1$	&	$-1$	&	1	&	$-1$	&	$1$	&	$1$
\\
11	&	$T_x^{-1}\T^{-1}T_x\T$			&	$1$	&	$1$	&	$1$	&	$1$	&	$-1$	&	1	&	$1$	&	$-1$	&	$-1$
\\
12	&	$T_y^{-1}\T^{-1}T_y\T$			&	$1$	&	$1$	&	$1$	&	$1$	&	$-1$	&	1	&	$1$	&	$-1$	&	$-1$
\\
13	&	$\T^2$						&	$-1$	&	$-1$	&	$-1$	&	$-1$	&	$-1$	&	1	&	$1$	&	$-1$	&	$-1$
\\\hline
\end{tabular}}
\caption{The columns labeled ``$s$PSG1-5," list the symmetry fractionalizations of the gapped, symmetric $\Zt$ spin liquids given in Table~\ref{tab:PSGs}. 
The corresponding bosonic symmetry fractionalization numbers are obtained by multiplying the $s$PSG numbers with the those given in the `vison' and `twist' columns. 
We see that $s$PSG5 corresponds to the $\Zt[0,0]$ state of Ref.~\onlinecite{YangWang}. No bosonic counterparts to $s$PSG1-4 are present in Ref.~\onlinecite{YangWang}.  }
\label{tab:SymmFrac}
\end{table}

In this section, we relate the gapped $\Zt$ spin liquids determined in the previous section to spin liquids obtained using Schwinger bosons by Yang and Wang \cite{YangWang} and Chatterjee {\it et al.\/} \cite{CQSS16}. 
Since these phases are gapped, they are completely defined via their ``symmetry fractionalization" \cite{EH13}.
Of the PSGs listed in Tables~\ref{tab:PSGs} and~\ref{tab:NematicPSGs}, we find that precisely one matches onto the spin liquid $A_b$, and one onto $B_b$ of Fig.~\ref{fig:2}a.
We begin by briefly reviewing this classification scheme in the context of $\Zt$ topological order.
The reader is referred to Ref.~\onlinecite{EH13} for more details.

One of the defining characteristics of topological order is the presence of anyonic excitations.
For the $\Zt$ case we consider here, there are two bosonic particles, typically denoted $e$ and $m$, which are mutually semionic: the wavefunction picks up a minus sign upon the adiabatic motion of an $e$ particle travelling around an $m$ particle. 
A bound state of an $e$ and $m$ is a fermionic excitation, $\vep\sim em$, and it also satisfies mutual semionic statistics with $e$ and $m$.  
We will frequently refer to the $m$ particle as the `vison' and the $e$ and $\vep$ particles as the bosonic and fermionic `spinon' respectively. 
These excitations carry $\Zt$ gauge charge and therefore must appear in pairs. 
Nonetheless, the $\Zt$ gauge field is gapped and these phases are deconfined, meaning that $e$, $m$, and $\vep$ particles may be very far from one another.

A comparison of these particles with the excitations in the Higgs phases implies that the fermionic spinons $\vep$ should be identified with the excitations of the field operator $X$.
In addition, in 2+1$d$ the Abrikosov vortices of the condensate are pointlike, and we associate these with the vison excitations $m$.
The remaining particle, the bosonic spinon $e$, is therefore described by a bound state of $X$ and the vortex.
In contrast, $\mathds{CP}^1$ is formulated in terms of the bosonic spinons. 
The vison is present as a vortex in the condensate as before, but now it is the fermionic spinon that is expressed as a bound state.

This representation of the degrees of freedom of a gapped $\Zt$ spin liquid provides a means to compare phases described using fermionic and bosonic ansatze. 
In a manner analogous to the classification of symmetries in terms of quantum numbers, these symmetry enriched topological phases can be classified 
by what is known as symmetry fractionalization numbers.
Independent of any formalism, suppose we create from the groundstate two $\vep$ (or $e$ or $m$) excitations and separate them so that they lie at very distant points $\v{r}$ and $\v{r}'$: $\Ket{\v{r},\v{r}'}$.
Since the rest of the system is indistinguishable from the groundstate, the action of an unbroken symmetry $G$ will exclusively affect these regions:
\eq{
G\Ket{\v{r},\v{r}'} &= \mathscr{G}_\vep(\br)\G_\vep(\br')\Ket{\br,\br'},
}
where $\mathscr{G}_\vep(\br)$ only has support in the region immediately surrounding $\br$.
As discussed in Sec.~\ref{sec:ModSyms}, the generators of a symmetry group satisfy certain commutation relations, and for the space group of the square lattice (plus time reversal), these relations are given in Eq.~\eqref{eqn:spaceGpRels} and below.
It follows that the action of any of these operations on \emph{all} wavefunctions must be equivalent to the identity.
For example, since $T_y^{-1}T_xT_yT_x^{-1}=\id$, it must map $\Ket{\br,\br'}$ to itself:
\eq{
\Ket{\br,\br'}&=T_y^{-1}T_xT_yT_x^{-1}\Ket{\br,\br'}.
}
In terms of the local symmetry operations, this becomes
\eq{
\Ket{\br,\br'}&=\sT_{\vep,y}^{-1}(\br)\sT_{\vep,x}(\br)\sT_{\vep,y}(\br)\sT_{\vep,x}^{-1}(\br)\cdot
\sT_{\vep,y}^{-1}(\br')\sT_{\vep,x}(\br')\sT_{\vep,y}(\br')\sT_{\vep,x}^{-1}(\br')
\Ket{\br,\br'}.
}
Since the transformations are localized at either $\br$ and $\br'$, they must be independent from one another and therefore constant.
However, because of the $\Zt$ gauge degree of freedom, $\zeta^\vep_{txty}=\sT_{\vep,y}^{-1}(\br)\sT_{\vep,x}(\br)\sT_{\vep,y}(\br)\sT_{\vep,x}^{-1}(\br)$ need not necessarily equal unity: the symmetry can be \emph{fractionalized} such that $\zeta_{txty}^\vep=-1$.
The value of $\zeta_{txty}^\vep$ will be consistent for every excitation of that species within a phase

It is not difficult to connect this to the PSG classification of the previous section. 
The PSG is the set gauge transformations required to preserve the form of the action following a symmetry transformation, as shown in Eq.~\eqref{eqn:PSGconstraint}.
Now, however, we present the PSG action solely in terms of operator which creates fermionic spinons, $X$:
\eq{
\mathscr{P}G:X\to U_GXV_G^\dag.
}
The same argument given above then requires that under the action of $T_y^{-1}T_xT_yT_x^{-1}$, $X$ is mapped to plus or minus itself:
\eq{
T_y^{-1}T_xT_yT_x^{-1}[X]=U_{tx}^\dag U_{ty}U_{xy}U_{ty}^\dag \,X\, V_{ty}V_{tx}^\dag V_{ty}^\dag V_{tx}=\pm X.
}
This factor is precisely the fractionalization number of $\vep$.
When time reversal is involved, this is modified to
\eq{
G^{-1}\T^{-1}G\T[X]&=
U_t^*\s^yU_G^*\s^yU_t^T U_G^\dag \,X\, V_GV_t^*\s^y V_G^T\s^yV_t^T,
\nt 
\T^2[X]&=U^*_t\s^y U_t \s^y\,X\, \s^yV_t^\dag \s^y V_t^T,
}
where the reality condition in Eq.~\eqref{eqn:RealityCondition} has been used.
Table~\ref{tab:symmGenNoDer} lists the numbers corresponding to each of the $s$PSG's in Table~\ref{tab:SymmFrac}.
(We note that the 7th group relation, $R^{-1}_{\pi/2}T_xR_{\pi/2}T_y=\id$, can be fixed by a gauge transformation on the relative sign of $V_{tx}$ and $V_{ty}$. 
In keeping with the convention of Ref.~\onlinecite{YangWang}, we require that the symmetry fractionalization number be $-1$ for the fermionic spinons.)

The argument also demonstrates a shortcoming of the PSG classification. 
While it immediately returns the symmetry fractionalization of the fermionic spinons, it provides no information regarding the symmetry fractionalization of the vison and bosonic spinon.
However, it fortunately turns out that the vison's fractionalization numbers are independent of the precise $\Zt$ spin liquid under study and can be obtained by examining a fully frustrated transverse-field Ising model \cite{RJSS91,MVSS99,Xu11,YangWang}. 
We quote these results in the column labeled ``vison" in Table~\ref{tab:SymmFrac}.

\subsubsection{Correspondence between fermionic and bosonic ansatze}
\label{sec:PSGcorres}
Comparing fermionic and bosonic ansatze may appear straightforward from this point: since $e\sim \vep\,m$, it seems reasonable to assume that the symmetry fractionalization of the bosonic spinon is obtained through a simple multiplication of the symmetry fractionalization numbers of the fermionic spinon and the vison.
However, the mutual statistics of $\vep$ and $m$ occasionally change this relation.
For instance, upon rotating $e$ by 360$^0$, $R_{\pi/2}^4$, either the vison will encircle the fermionic spinon or vice versa. 
In either case, an extra factor of $-1$ must be taken into account. 
These additional multiplication factors were worked out in Ref.~\onlinecite{YangWang}, and we quote them under the column labeled ``twist" in Table~\ref{tab:SymmFrac}.

The comparison with the bosonic symmetry fractionalization allows us to identify $s$PSG5 with $\Zt[0,0]$, showing that $A_f=A_b$ as promised. 
We do not find fermionic counterparts to the remaining four spin liquids in Ref.~\onlinecite{YangWang}.

Using a slightly altered set of commutation relations to account for the symmetry breaking, the exact same analysis can be performed for the nematic spin liquids.
These symmtery fractionalization numbers are shown in Table~\ref{tab:NematicPSGs}, and,
as claimed, by comparing with the analysis of Ref.~\cite{CQSS16} we positively identify $B_f$ ($n$PSG7) with the Ising-nematic $\Zt$ spin liquid $B_b$.

Finally, the equivalence of $A_f$ and $B_f$ with $A_b$ and $B_b$ indicates the equivalence of $C_f$ and $C_b$. 
In Appendix~\ref{app:CLsymFrac}, we provide additional verification of this result using the symmetry fractionalization technique.

\subsection{Lattice classification of fermionic PSG's}

The data compiled in Table~\ref{tab:SymmFrac} can also be used to compare the phases we find against fermionic spin liquids described on the lattice.
In Appendix~\ref{app:WenClassification}, we review Wen's classification scheme \cite{Wen02} and identify the lattice PSG's corresponding to the two U(1) spin liquids as well as the five symmetric $\Zt$ spin liquids.
This classification is useful since it allows us to express the phases we've studied on the lattice without having to reverse engineer the bilinears.

We identify the gapped U(1) spin liquid, $D_f$, with $\mathrm{U1C}n0n1$ and the gapless U(1) spin liquid ($\Braket{\Phi_1}\neq0$) with $\mathrm{U1B}x11n$. 
The lattice PSG's corresponding to the five symmetric $\Zt$ spin liquids we obtained are shown in Table~\ref{tab:LatticeNames}.

Both $s$PSG1 and $s$PSG5 seemingly correspond to multiple lattice PSG's. 
However, in Appendix~\ref{app:Z2latticePSGcorr}, we prove that while the spin liquids have the same symmetry fractionalizations, of the two shown, only one of each pair actually corresponds to the spin liquids we consider.
In the case of $s$PSG5, it is not difficult to show that $\mathrm{Z2B}xx2z$ is always gapless, immediately ruling it out as a description of the gapped phase $A_f$.
Further, we show that $\mathrm{Z2B}xx2z$ is not proximate to either the gapped or gapless U(1) spin liquids U1C$n0n1$ and U1B$x11n$.
Similarly, we find that $\mathrm{Z2B}xx23$ is not proximate to U1C$n0n1$, leaving $\mathrm{Z2B}xx13$ as the sole realizable lattice PSG capable of reproducing $s$PSG1.

These statements can be verified explicitly by comparing our continuum theory with mean field Hamiltonians on the lattice which have been constructed using only information provided by the lattice PSG. 
In Appendix~\ref{app:LatticeHamiltonians} we study the lattice Hamilonians for the gapped and gapless U(1) spin liquids, as well as $A_f$.
We find that a low-energy expansion of the mean field Hamiltonian describing U1C$n0n1$ corresponds to adding $\tr\(\s^x\bX \m^y X\)$ to the $\pi$-flux Hamiltonian as expected, but that no analogous statement can be made for either U1B$x11n$ or Z2B$xx1z$.
In particular, we demonstrate that no mean field ansatz on the lattice can realize the U(1) spin liquid U1B$x11n$.
This should not be too surprising as the continuum realization of this phase is the product of condensing $\Phi_1$ when coupled to $\tr\(\s^a\bX \ptl_0 X\)$, the time component of a vector.
This description is manifestly dependent on the presence of temporal fluctuations in contrast to the purely static mean field analysis. 

Conversely, a lattice Hamiltonian describing the $\Zt$ phase $A_f$ does exist.
However, upon expanding the resulting Hamiltonian about its Dirac cones, the hopping term which breaks the U(1) symmetry down to $\Zt$ appears to arise from coupling $\tr\(\s^a\bX \m^y\ptl_x\ptl_y\[\ptl_x^2-\ptl_y^2\]X\)$ to a condensed Higgs field. 
We can see why this may be the case by observing how symmetries act on $\Xi=\tr\(\s^a\bX \m^y\ptl_x\ptl_y\[\ptl_x^2-\ptl_y^2\]X\)$:
\eq{
\T[\Xi]&=-\Xi,
&
P_{x,y}[\Xi]&=-\Xi,
&
T_{x,y}[\Xi]&=-\Xi,
&
R_{\pi/2}[\Xi]&=\Xi.
}
It follows that a Higgs field $\Phi'_1$ which couples to $\Xi$ may have a non-zero expectation value in the $A_f$ phase provided it is perpendicular in colour space to both $\Phi$ and $\Phi_1$.
That is, supposing $\Braket{\Phi^x}\neq0$ and $\Braket{\Phi_1^z}\neq0$, having $\Braket{{\Phi'_1}^z}\neq0$  will not break any of the symmetries.

It can also be shown that the Ising-nematic spin liquid, $B_f$, is not `anomalous' in the manner just discussed.
\begin{table}
\centering
\begin{tabular}{| c || c | c | c | c | c |}
\hlinewd{.8pt}
    &   $s$PSG1  &   $s$PSG2 &   $s$PSG3 &   $s$PSG4 &   $s$PSG5 ($A_f$)
    \\\hlinewd{.8pt}\hline
\multirow{2}{*}{Lattice PSG}    &
$\mathrm{Z2B}xx13$  &   \multirow{2}{*}{$\mathrm{Z2B}xx03$}  &  \multirow{2}{*}{$\mathrm{Z2B}0013$} &   \multirow{2}{*}{$\mathrm{Z2B}0001$}  &   
\cellcolor{Yellow!40} $\mathrm{Z2B}xx1z$  \\
    &  $\mathit{Z2Bxx23}$  &   &   &   & $\mathit{Z2Bxx2z}$ 
\\\hlinewd{.8pt}
\end{tabular}
\caption{Spin liquids according to the labeling scheme given in Ref.~\onlinecite{Wen02} and reviewed in Appendix~\ref{app:Z2latticePSGcorr}.
All of the spin liquids listed are found to be proximate to the $\pi$-flux phase SU2$n0$ though not necessarily U1C$n0n1$.
While the symmetry fractionalization of $s$PSG1 and $s$PSG5 corresponds to multiple fermionic PSG's, the two which have been italicized ($\mathrm{Z2B}xx23$ and $\mathrm{Z2B}xx2z$) are not proximate to U1C$n0n1$ and therefore cannot represent the Higgs phases we obtain (see Appendix~\ref{app:Z2latticePSGcorr}).}
\label{tab:LatticeNames}
\end{table}

\begin{table}
\centering
{\footnotesize
\begin{tabular}{|r |l || r | r  ||*{6}{r|}>{\columncolor{SkyBlue!20}}r|*{3}{r|}}
\hline
& \multirow{2}{*}{Group relations}	
& \multicolumn{1}{c|}{\multirow{2}{*}{v}} 
& \multicolumn{1}{c||}{\multirow{2}{*}{t}}
& \multicolumn{10}{c|}{$n$PSG$x$}  
\\\hhline{*{4}{|~}*{10}{|-}}
&	&   &   & 1	&	2	&	3	&	4	&	5	&	6   &   7	&	8   &   9   &   10		 \\\hline\hline
1	&	$T_y^{-1}T_xT_yT_x^{-1}$	
&  $-1$  &	 $1$  
& $1$  & $1$  &   $1$  &  $-1$  &    $-1$  &    $-1$  &    $-1$  &    $1$   &   $1$ &   $1$
\\\hline
2	&	$P_x^{-1}T_xP_xT_x$	
&   $1$ &   $1$ 
&	$-1$    &   $1$  &    $-1$  &    $1$  &   $-1$  &    $1$  &    $1$  &     $1$  &   $1$  &   $-1$
\\\hline
3	&	$P_y^{-1}T_xP_yT_x^{-1}$	
&   $-1$    &   $1$
&   $1$	&	$1$	&	$-1$	&	$-1$	&	$-1$	&	$1$	&	$-1$	&	$-1$  &   $1$  &   $-1$
\\\hline
4	&	$P_x^{-1}T_yP_xT_y^{-1}$	
&   $-1$  & $1$ 
&	$-1$	&	$1$	&	$-1$	&	$1$	&	$-1$	&	$1$	&	$1$	&	$1$	  &   $1$  &   $-1$
\\\hline
5	&	$P_y^{-1}T_yP_yT_y$	
&   $1$    &   $1$
&	$1$	&	$1$	&	$-1$	&	$-1$	&	$-1$	&	$1$	&	$-1$	&	$-1$  &   $1$  &   $-1$
\\\hline
6	&	$P_x^2$	
&   $1$    &    $-1$
&	$-1$    &	$1$	&	$-1$	&	$1$	&	$-1$	&	$1$	&	$1$	&	$1$  &   $1$  &   $-1$
\\\hline
7	&	$P_y^2$	
&   $1$   & $-1$
&	$1$	&	$1$	&	$-1$	&	$-1$	&	$-1$	&	$1$	&	$-1$	&	$-1$  &   $1$  &   $-1$
\\\hline
8	&	$P_x^{-1}P_y^{-1}P_xP_y$	
&   $-1$    &    $-1$
&	$1$	&	$1$	&	$1$	&	$1$	&	$1$	&	$1$	&	$1$	&	$1$  &   $1$  &   $1$
\\\hline
9	&	$T_x^{-1}\T^{-1}T_x\T$	
&   $1$   & $1$
&	$-1$    &	$-1$	&	$1$	&	$-1$	&	$1$	&	$1$	&	$1$	&	$1$  &   $1$  &   $-1$
\\\hline
10	&	$T_y^{-1}\T^{-1}T_y\T$	
&   $1$  &  $1$
&	$1$    &	$1$	&	$-1$	&	$-1$	&	$1$	&	$1$	&	$1$	&	$-1$	  &   $-1$  &   $1$
\\\hline
11	&	$P_x^{-1}\T^{-1}P_x\T$	
&   $1$  &  $-1$
&	$-1$    &	$1$	&	$-1$	&	$1$	&	$-1$	&	$1$	&	$1$	&	$1$  &   $1$  &   $-1$
\\\hline
12	&	$P_y^{-1}\T^{-1}P_y\T$	
&   $1$    &  $-1$
&	$1$ 	&	$1$	&	$-1$	&	$-1$	&	$-1$	&	$1$	&	$-1$	&	$-1$  &   $1$  &   $-1$
\\\hline
13	&	$\T^2$	
&   $1$    &  $1$
&	$-1$&	$-1$	&	$-1$	&	$-1$	&	$-1$	&	$-1$	&	$-1$	&	$-1$  &   $-1$  &   $-1$
\\\hline
\end{tabular}}
\caption{Symmetry fractionalization of nematic PSG's ($n$PSG's) for spin liquids listed in Table~\ref{tab:NematicPSGs}. $n$PSG7$x$ (highlighted in blue) corresponds to the fermionic PSG determined in \cite{CQSS16}.
The columns labelled `v' and `t' list the vison fractionalization numbers and the twist factors respectively.
}
\label{tab:nPSGsymFrac}
\end{table}

\subsection{Identification of U(1) spin liquid}\label{sec:BerryPhase}

The arguments which allowed us to compare $\Zt$ spin liquids expressed using bosonic and fermionic spinons breaks down in the presence of gapless degrees of freedom.
In both cases, these phases are unstable to the proliferation of monopoles, and their true ground states will break any symmetries under which the monopoles transform nontrivially. 
In order to ensure that $D_f$ actually corresponds to the massive phase of the $\mathds{CP}^1$ theory, $D_b$, we verify that the two spin liquids share the same fate and ultimately realize a VBS.
We approach this problem from two perspectives.
We first follow the method outlined in Ref.~\onlinecite{fu11} and determine which bilinear operators \emph{respond} to a weakly varying flux and, consequently, the monopoles' presence.
We complement this analysis by calculating the Berry phase of the monopole in a certain limit and show that it agrees with the analogous calculation performed using Schwinger bosons in Ref.~\onlinecite{NRSS90}.

\subsubsection{Flux Response}

The effective Lagrangian describing $D_f$ is
\eq{
\mathcal{L}_{\mathrm{U}(1)}&=i\tr\( \bX\g^\m \[\ptl_\mu X+i X \s^xa^x_\m\]\)+\lam_2 \Braket{\Phi^x}\tr\(\s^x\bX \m^y X\).
}
Because both $a_\m^y$ and $a_\m^z$ are gapped, only $a_\m^x$ is included in $\L_{\mathrm{U}(1)}$.
In what follows, we drop the `$x$' index, taking $a_\m^x\to a_\m$ (this $a_\m$ should not be confused with the gauge field of the original SU(2) gauge field).
Finally, at this point in the discussion, it is more convenient to express the Lagrangian in terms of Dirac spinors.
Using Eq.~\eqref{eqn:DiracMajRelation}, we find
\eq{
\L_\mathrm{U(1)}&=\bar{\psi}i\g^\m\(\ptl_\m-ia_\m\s^x\)\psi+m\bpsi \s^x\m^y \psi 
}
where $m=\lam_2\Braket{{\Phi}^x}$.

In the context of a U(1) gauge theory, a monopole is a topologically nontrivial field configuration of $a_\m$. 
In imaginary time, this configuration corresponds exactly to a (stationary) Dirac monopole in 3+1$d$ electromagnetism.
However, instead of behaving as a particle, in 2+1$d$ the monopole is actually an instanton: it describes tunneling between different vacua or topological sectors labeled by their total flux, $\int dS_\m\ep^{\m\n\lam}\ptl_\n a_\lam=2\pi n$ where $n$ is an integer. 
This number is zero in the deconfined phase of the gauge theory whereas it fluctuates and ceases to take a definite value once the monopoles proliferate.

A complete treatment of the monopole proceeds by first expanding the gauge field into a classical background piece $A_\m$ and a quantum fluctuation piece $\tilde{a}_\m$, $a_\m=A_\m+\tilde{a}_\m$, and quantizing the theory about this background.
Because the monopole background breaks translational symmetry, this is quite an involved calculation which we will not perform.
Instead, we investigate the impact a non-zero flux has on the other operators of the theory.
That is, we assume that the classical monopole configuration described by $A_\m$ varies very slowly and, through linear response, determine which operators, $\O$, the flux couples to at leading order:
$\Braket{\O(x)}=\int d^3x'\, \chi_\O^\m(x,x')A_\m(x')$ where $\chi_\O^\m=\Braket{\O(x)\bar{\psi}\g^\m\s^x\psi(x')}$.
This calculation is outlined in Appendix~\ref{app:fluxResponse} and at low energies yields
\eq{\label{eqn:fluxLinResp}
\Braket{\bpsi \g^\m\m^y \psi}={1\o\pi}\ep^{\m\a\b}\ptl_\a A_\b.
}
Consequently, whenever there is a net flux, $\int d^2x \(\ptl_x A_y-\ptl_y A_x\)\neq0$, we expect $\Braket{\bpsi\g^0\m^y\psi}\neq0$ as well. 
This allows us to identify $\bpsi \g^\m\m^y\psi$ with the topological current.
The topological charge is then obtained by integrating the zeroth component of the current over space:
\eq{
Q={1\o2}\int d^2r\,\bar{\psi}\g^0\m^y\psi.
}
The factor of $1/2$ is chosen to ensure that $Q$ is always an integer, as follows from Eq.~\eqref{eqn:fluxLinResp}.

A conserved charge is the generator of the associated symmetry, meaning that $Q$ generates the flux conservation symmetry.
However, this operator should be familiar from Sec.~\ref{sec:Higgsf} where it was observed to be the generator of the U(1) VBS symmetry.
This can be confirmed by checking that
\eq{
\[ Q,V_x\]&=iV_y,
&
\[Q,V_y\]&=-iV_x.
}
where $V_x={1\o2}\bpsi\m^x\psi$, $V_y={1\o2}\bpsi\m^z\psi$. 
It follows that $Q$ is conjugate to the VBS order parameters.

When the gapped U(1) gauge theory confines, the monopole proliferation induces large fluctuations in $Q$. 
This in turn suppresses the fluctuations of the operators conjugate to $Q$, ultimately resulting in long range order.
We conclude from the analysis above that the proliferation of monopoles triggers the condensation of one of the VBS order parameters, proving that $D_f$ is unstable to a VBS and therefore equivalent to $D_b$.
This mechanism should be contrasted with the scenario outlined in Sec.~\ref{sec:duality} where VBS order was achieved by tuning $\m>0$ in Eq.~\eqref{eqn:AnisotropicOp}.

\subsubsection{Berry phase}\label{sec:BerryPhaseSubSec}

A separate argument for the identification of the U(1) spin liquid proceeds by a computation of the monopole Berry phase, along the lines of the original arguments using the semiclassical quantization of the antiferromagnet \cite{Haldane88}, or the
Schwinger boson theory of the U(1) spin liquid \cite{NRSS90}.
Here, this argument starts from a lattice Hamiltonian for the U1C$n0n1$ U(1) spin liquid, which we obtain from Eq.~(\ref{U1MF})
for a generic direction of the Higgs field $\Phi$
\eq{
H &=-\sum_\vi\bigg\{
{i\a}\( \psi^\dag_\vi\psi_{\vi+\hx}+(-)^{i_x}\psi_\vi^\dag\psi_{\vi+\hy}+h.c.\)
+ \Phi^a (-)^{i_x+i_y}
\b\(\psi_\vi^\dag \t^a\psi_{\vi+2\hx}+\psi_\vi^\dag\t^a\psi_{\vi+2\hy}+h.c.\)
\nt
&\quad
-\Phi^a a_0 (-)^{i_x+i_y}\psi^\dag_\vi\t^a\psi_\vi
\bigg\}. \label{U1MFa}
}
We are interested in saddle-points of the associated action where the $\Phi^a$ Higgs field, and the associated SU(2) gauge field (not written explicitly in Eq.~(\ref{U1MFa})) take the configuration of
't~Hooft-Polyakov monopoles \cite{tH74,Polyakov74} in 2+1 dimensional spacetime. After obtaining such saddle points, we have to compute the fermion determinant in such a background, and the phase of this determinant will yield the needed monopole Berry phase. This is clearly a demanding computation, and we will not attempt to carry it out in any generality. However, assuming the topological invariance of the needed quantized phase, we can compute it by distorting the saddle point Lagrangian, without closing the fermion gap, to a regime where the phase is easily computable. Specifically, consider the limit where the parameter $a_0$ in Eq.~(\ref{U1MFa}) is much larger than all other parameters, including $\alpha$ and $\beta$. For the 't Hooft-Polyakov monopole at the origin of spacetime, 
$\Phi^a \sim \hat{r}^a$, where $\hat{r}^a$ is a unit radial vector in spacetime. Ignoring all but the $a_0$ term in Eq.~(\ref{U1MFa}), we then have to compute the Berry phases of 
single fermions, each localized on a single site, in the presence of a staggered field $\propto \Phi^a$. However, this Berry phase
is precisely that computed by Haldane \cite{Haldane88}; in his case the staggered field was the antiferromagnetic order parameter which acts in the spin SU(2) space (in contrast to the staggered field in the gauge SU(2) space in our case), and the Berry phase arose from that of a quantized $S=1/2$ spin. As the Berry phase of a spin $1/2$ localized fermion is equal to the spin Berry phase, we conclude that the 't~Hooft-Polyakov monopole Berry phase is equal to that obtained by Haldane \cite{Haldane88} for $S=1/2$. Therefore, the monopole Berry phases in the fermionic spinon U(1) spin liquid U1C$n0n1$
are equal to those of the U(1) spin liquid of the bosonic
$\mathbb{CP}^1$ theory of the square lattice antiferromagnet \cite{AA89,SJ90}.



\section{Conclusions}
\label{sec:conc}

Two distinct classes of $2+1$ dimensional 
fermion-boson dualities have recently seen
much discussion in the literature.

One class concerns gapped $\mathbb{Z}_2$ spin liquids which have
both bosonic and fermionic spinon excitations. Binding with a vison transmutes a spinon from a boson
to a fermion, or vice versa \cite{RC89}, and this allows one to map between $\mathbb{Z}_2$ spin liquids obtained
in mean-field theory using fermionic or bosonic ansatze. Specific examples of such dualities
have been described on a variety of lattices \cite{YY12,EH13,LCV14,QiFu15,ZLV15,ZMQ15,Lu15,YangWang,CQSS16},
and our results for such dualities appear in Figs.~\ref{fig:1} and~\ref{fig:2}.
We described the dualities between the bosonic $\mathbb{Z}_2$ spin liquids
$A_b$, $B_b$, $C_b$ and the fermionic spin liquids $A_f$, $B_f$, $C_f$
respectively. The first two of these dualities have been obtained earlier \cite{YangWang,CQSS16}, but we obtained all three in a unified manner with
reference to continuum theories. 

The second class of dualities concern conformal gauge 
theories with fermionic and bosonic matter \cite{SSWW16,KT16,Murugan16,MVX17,Kachru16,Raghu17}. 
Most relevant to our considerations
is the proposed duality \cite{Wang17} between the critical 
bosonic $\mathbb{CP}^1$ U(1) gauge theory, and fermionic SU(2) QCD$_3$
with $N_f=2$ flavors of Dirac fermions. 

Among our results was a demonstration of the compatibility between
these two classes of dualities. We Higgsed the critical bosonic $\mathbb{CP}^1$ and
fermionic QCD$_3$ theories, and found nontrivial consistency between the gapped $\mathbb{Z}_2$
spin liquids so obtained. We also obtained a fermionic counterpart to the U(1) spin liquid 
with gapped bosonic spinons
on the square lattice originally obtained by Arovas and Auerbach \cite{AA89} (which is equivalent to the gapped $z_\alpha$ phase of the $\mathbb{CP}^1$ theory \cite{NRSS89,NRSS90}):
the U(1) spin liquid with gapped fermionic spinons was identified as U1C$n0n1$ (in Wen's notation). Both the bosonic and fermionic U(1) spin liquids are eventually unstable to monopole proliferation, confinement, and VBS order, and have identical
monopole Berry phases (as shown in Section~\ref{sec:BerryPhase}).

Our analysis also led us to propose new fermion-boson 
dualities between multi-critical theories.
One example is the duality between
({\it i\/}) the U(1) gauge theory in Eq.~(\ref{b1}) with two unit charge bosons $z_\alpha$, a doubly charged Higgs field $P$, and the masses of both fields tuned to criticality; and ({\it ii}) the SU(2) gauge theory in Eq.~(\ref{f2}) with $N_f=2$ massless fundamental Dirac fermions $\psi$, and two adjoint Higgs fields $\Phi$, $\Phi_1$, and the masses of both Higgs 
field tuned to criticality.

The fermionic approach to square lattice spin liquids \cite{AM88,Affleck88,Wen02} yields a variety of critical
spin liquids coupled to U(1) and SU(2) gauge fields. Two examples are in Fig.~\ref{fig:1}b, 
the states labeled by Wen as U1B$x11n$ and SU2B$n0$. The results of Wang {\it et al.\/} \cite{Wang17} indicate that the SU(2) critical state SU2B$n0$ cannot appear as an extended
critical phase in a square lattice antiferromagnet, and it is only realized as a critical
point between the N\'eel and VBS states. From our comparison of Fig.~\ref{fig:1}b
and Fig.~\ref{fig:1}a, we obtain evidence that the critical U(1) spin liquid
U1B$x11n$ also cannot be realized as an extended phase on the square lattice: it is unstable
to the appearance of canted antiferromagnetic order.


\section*{Acknowledgements}

We thank Max Metlitski and Cenke Xu for valuable discussions.
This research was supported by the National Science Foundation under Grant No. NSF PHY-1125915 at KITP,
and under Grant No. DMR-1360789 at Harvard. Research at Perimeter Institute is supported by the Government of Canada through Industry Canada and by the Province of Ontario through the Ministry of Research and Innovation. SS also acknowledges support from Cenovus Energy at Perimeter Institute. 

\appendix

\section{Spin liquids with projective spin symmetry}
\label{app:ProjSpinSymm}
We expand upon our assertion in Sec.~\ref{sec:symmliq} that a fully symmetric, gapped spin liquid cannot be obtained through the condensation of a Higgs field $\Phi$ coupling to a bilinear which transforms in a nontrivial manner under the SU(2) spin symmetry. 
As discussed, in order for the resulting spin liquid to have a spin gap, $\Phi$ must couple to one of the operators in Tab.~\ref{tab:symmGenNoDer}.
We start by studying $N^{ab}=\tr\(\s^a \bX \s^b X\)$ and couple it to a Higgs field as $\sum_{a,b}\Phi^{ab}N^{ba}=\tilde{\tr}\(\Phi N\)$, where `$\tilde{\tr}$' refers to a trace over the spin and colour \emph{vector} labels (as opposed to the usual trace `$\tr$' over spin and colour \emph{spinor} indices).
In the Higgs phase, we write $\bar{\Phi}=\Braket{\Phi}\neq0$.

Naturally, having the Higgs couple to $N^{ab}$ implies that spin symmetry is realized projectively in the condensed phase, if at all.
We associate SO(3) matrices to both the SU(2) gauge and spin transformations.
That is, instead of studying the action of gauge and spin transformations $U_g$ and $U_s$, we consider matrices $Q,R\in\mathrm{SO}(3)$ such that
\eq{
\mathrm{SU}(2)_s: N^{ab} &\to \tr\(\s^a\bX U^\dag_s \s^b U_s X\) = N^{ac}\(R^T\)^{cb},
\nt
\mathrm{SU}(2)_g: N^{ab} &\to \tr\( U_g\s^aU^\dag_g\bX \s^b X\) = Q^{ac}N^{cb}.
}
Under a projective spin transformation, 
\eq{
\mathscr{P}\mathrm{SU}(2)_s: \tilde{\tr}\(\bar{\Phi} N\)\to \tilde{\tr}\(\bar{\Phi} Q N R^T\)=\tilde{\tr}\(\bar{\Phi} N\),
}
implying that
$Q=\bar{\Phi}^{-1}R$.
The requirement that $Q\in\mathrm{SO}(3)$ implies that $\bar{\Phi}\in\mathrm{SO}(3)$ as well, for example $\bar{\Phi}^{ab}=\abs{\Phi}\d^{ab}$.

The obstruction to forming a fully symmetric spin liquid is then apparent. 
Since $N^{ab}\to -N^{ab}$ under $\T$, $P_x$ and $P_y$, the equivalence of the original and symmetry transformed states requires that $\bar{\Phi}$ be gauge equivalent to $-\bar{\Phi}$.
This in only possible if $Q_{t,px,py}=-\id\not\in\mathrm{SO}(3).$

These considerations apply equally to $\tr\(\s^a\bX \m^{x,z}\s^b X\)$ as indicated in Section~\ref{sec:symmliq}.

\section{Wen's Lattice PSG Classification Scheme}
\label{app:WenClassification}

In this appendix, we relate our results to the spin liquid classification scheme proposed in Ref.~\onlinecite{Wen02} by Wen.
We begin by reviewing his conventions and formalism before explaining what it means for two spin liquids to be ``proximate" in this language.
We then discuss how we determined that the gapped and gapless U(1) spin liquids in Fig.~\ref{fig:1}b correspond to $\mathrm{U1C}n0n1$ and $\mathrm{U1B}x11n$ respectively.
We subsequently consider the $\Zt$ $s$PSG's and explain
how the identification in Table~\ref{tab:LatticeNames} was obtained.

We note that frequent reference will be made to information that is only present in the arXiv version of Ref.~\cite{Wen02}.

\subsection{Conventions and formalism}

Here, we briefly review the spin liquid classification scheme proposed in Ref.~\onlinecite{Wen02}; for a complete discussion the reader is referred to the original paper.
In keeping with these conventions, we express the mean field Hamiltonian of Eq.~\eqref{eqn:MFHamiltonian} in terms of fermions $\psi=\(\psi_1,\psi_2\)^T=\(f_\uparrow,f_\downarrow^\dag\)^T$.
The mean field ansatz is written in terms of the matrix
\eq{
u_{\vi\vj}&=
{3\o8}J\begin{pmatrix}
\a_{\vi\vj}^\dag   &   \b_{\vi\vj} \\
\b_{\vi\vj} &   -\a_{\vi\vj}    
\end{pmatrix}
=u_{\vj\vi}^\dag.
}
The average constraint in Eq.~\eqref{eqn:aveSU2GaugeConstraint} become s
\eq{\label{eqn:aveSU2GaugeConstraint3}
\Braket{\psi_\vi^\dag \t^\ell \psi_\vi}=0
}
where $\t^\ell$ are Pauli matrices (with $\t^0=\id$) and the mean field Hamiltonian can then be written
\eq{\label{eqn:MFgeneral}
H_\mathrm{MF}&=
\sum_{\Braket{ij}}
\bigg[
{4\o3J_{\vi\vj}}\tr\(u^\dag_{\vi\vj}u_{\vi\vj}\)-\(\psi_\vi^\dag u_{\vi\vj}\psi_\vj +h.c.\)\bigg]+\sum_\vi a_0^\ell \psi_\vi^\dag \t^\ell \psi_\vi .
}
Here, $u_{\vi\vj}$ is the analogue to $\a_{\vi\vj},\b^a_{\vi\vj}$ (when $\g_{\vi\vj}\neq0$ the spin symmetry is realized projectively, a possibility this formalism does not take into account \cite{Chen2012}). 
$a_0^\ell$ are Lagrange multipliers enforcing the constraint in Eq.~\eqref{eqn:aveSU2GaugeConstraint3}.
In order for $H_\mathrm{MF}$ to preserve spin, we must choose $iu_{\vi\vj}\in \mathrm{SU}(2)$.
Finally, the SU(2) gauge symmetry acts on the $\psi$ fermions and ansatz as
\eq{
\psi_\vi&\to W(\vi)\psi_\vi,
&
u_{\vi\vj}\to W(\vi)u_{\vi\vj}W^\dag(\vj).
}

The projective symmetry group in this context is expressed as the invariance of the ansatz $u_{\vi\vj}$ under the joint action of a symmetry transformation $G$ and a gauge transformation $W_G$: Eq.~\eq{\label{eqn:AnsatzInvariance}
W_GG\[u_{\vi\vj}\]=u_{\vi\vj}
}
where
\eq{
G[u_{\vi\vj}]&=u_{G(\vi),G(\vj)}
&
W_G\[u_{\vi\vj}\]&=W_G(\vi)u_{\vi\vj}W_G^\dag(\vj),
&
W_G(\vi)\in \mathrm{SU}(2).
}
Here, we have assumed that $G$ is a space group operation; for time reversal, we have $\T[u_{\vi\vj}]=-u_{\vi\vj}$.
The invariant gauge group (IGG) is the set of gauge transformations which do not alter the ansatz,
\eq{
\gW&=\{W(\vi)\, |\, W(\vi)u_{\vi\vj}W(\vj)^\dag,W(\vi)\in\mathrm{SU}(2) \},
}
and, therefore, $\gW$ can either be SU(2), U(1), or $\Zt$.
In the main body of the text, this is what we simply refer to as the gauge group or, sometimes in a Higgs phase, the ``residual gauge group." 

In order to make use of the symmetry fractionalization technique, we translate the commutation relations in Eq.~\eqref{eqn:spaceGpRels} and below to the lattice case:
 \eq{\label{eqn:LatticeGpRels}
1.&
&
W^{-1}_{ty}(i_x,i_y+1)W_{tx}(i_x,i_y+1)W_{ty}(i_x-1,i_y+1)W_{tx}^{-1}(i_x,i_y)&\in \gW
\nt 
2.&
&
W_{py}^{-1}(i_x,-i_y)W_{tx}(i_x,-i_y)W_{py}(i_x-1,-i_y)W_{tx}^{-1}(i_x,i_y)&\in \gW
\nt 
3.&
&
W_{py}^{-1}(i_x,-i_y)W_{ty}(i_x,-i_y)W_{py}(i_x,-i_y-1)W_{ty}(i_x,i_y+1)&\in \gW
\nt 
4.&
&
W_{py}(i_x,i_y)W_{py}(i_x,-i_y)&\in \gW
\nt 
5.&
&
W_{py}^{-1}(i_x,-i_y)W_r(i_x,-i_y)W_{py}(-i_y,-i_x)W_r(-i_y,i_x)&\in \gW
\nt 
6.&
&
W_r(i_x,i_y)W_r(i_y,-i_x)W_r(-i_x,i_y)W_r(-i_y,i_x)&\in \gW
\nt 
7.&
&
W_r^{-1}(-i_y,i_x)W_{tx}(-i_y,i_x)W_r(-i_y-1,i_x)W_{ty}(i_x,i_y+1)&\in \gW
\nt 
8.&
&
W_r^{-1}(-i_y,i_x)W_{ty}(-i_y,i_x)W_r(-i_y,i_x-1)W_{tx}^{-1}(i_x,i_y)&\in \gW
\nt 
9.&
&
W_t^{-1}(i_x,i_y)W_r^{-1}(-i_y,i_x)W_t(-i_y,i_x)W_r(-i_y,i_x)&\in \gW
\nt
10.&
&
W_t^{-1}(i_x,i_y)W_{py}^{-1}(i_x,-i_y)W_t(i_x,-i_y)W_{py}(i_x,-i_y)&\in \gW
\nt
11.&
&
W_t^{-1}(i_x,i_y)W_{tx}^{-1}(i_x+1,i_y)W_t(i_x+1,i_y)W_{tx}(i_x+1,i_y)&\in \gW
\nt 
12.&
&
W_t^{-1}(i_x,i_y)W_{ty}^{-1}(i_x,i_y+1)W_t(i_x,i_y+1)W_{ty}(i_x,i_y+1)&\in \gW
\nt 
13.&
&
W_t(i_x,i_y)W_t(i_x,i_y)&\in\gW 
}

\subsection{SU(2) spin liquid classification}
We presented the mean field ansatz of the $\pi$-flux phase in Sec.~\ref{sec:PiFluxAnsatz}.
In Wen's notation, it corresponds to the spin liquid SU2B$n0$, and consequently has the following PSG:
\eq{\label{eqn:SU2latPSG}
W_{tx}(\vi)&=(-)^{i_y}g_{tx},
&
W_{px}(\vi)&=(-)^{i_x}g_{px},
&
W_{pxy}(\vi)&=(-)^{i_xi_y}g_{pxy},
\nt
W_{ty}(\vi)&=g_{py},
&
W_{py}(\vi)&=(-)^{i_y}g_{py},
&
W_t(\vi)&=(-)^{i_x+i_y}g_t,
}
where $g_\xi\in \mathrm{SU}(2)$, $\xi=tx,ty,px,py,pxy,t$.
All PSG's proximate to SU2B$n0$ can be obtained by fixing the values of the $g_\xi$ to a specific element in SU(2) (the PSG's are only defined modulo the IGG).
In Appendix B of Ref.~\onlinecite{Wen02}, Wen enumerates which U(1) and $\Zt$ PSG's are proximate to SU2B$n0$.
All of the phases we consider must be identified with one of these options.

\subsection{U(1) spin liquid classification} \label{sec:U1id}
\begin{table}
\centering
{\footnotesize
\begin{tabular}{ | r | l || l | l |}
\hlinewd{.8pt}
	&	 \multicolumn{1}{c||}{Group relations}	& \multicolumn{1}{c|}{Gapped ($D_f$)}	&	 \multicolumn{1}{c|}{Gapless}		\\\hlinewd{.8pt}\hline	
1	&	$T_y^{-1}T_xT_yT_x^{-1}$ 		&	
$-e^{-2i\(\th_{tx}-\th_{ty}\)\s^z}$	&	
$-\id$	
\\		
2	&	$P_y^{-1}T_xP_yT_x^{-1}$		&	$\ph{-}e^{2i\th_{py}\s^z}$		&	
    $-e^{2i\th_{tx}\s^z}$	
\\
3	&	$P_y^{-1}T_yP_yT_y$			&	$\ph{-}e^{2i\th_{py}\s^z}$		&
    $-\id$			
\\
4	&	$P_y^2$						&	$\ph{-}e^{-2i\th_{py}\s^z}$		&	$-\id$
\\
5	&	$P_y^{-1}R_{\pi/2}P_yR_{\pi/2}$	&	$\ph{-}e^{-2i\th_{r}\s^z}$		&	$\ph{-}\id$
\\
6	&	$R_{\pi/2}^4$					&	$\ph{-}e^{-4i\th_{r}\s^z}$		&	
$\ph{-}e^{-4i\th_r\s^z}$
\\
7	&	$R_{\pi/2}^{-1}T_xR_{\pi/2}T_y$	&	$\ph{-}e^{2i\th_r\s^z+i\(\th_{tx}-\th_{ty}\)\s^z}$	&$-e^{-i\(\th_{tx}+\th_{ty}\)\s^z}$
\\
8	&	$R_{\pi/2}^{-1}T_yR_{\pi/2}T_x^{-1}$	&	$\ph{-}e^{2i\th_r\s^z-i\(\th_{tx}-\th_{ty}\)\s^z}$	&$\ph{-}e^{i\(\th_{tx}-\th_{ty}\)\s^z}$
\\
9	&	$R_{\pi/2}^{-1}\T^{-1}R_{\pi/2}\T$	&	$\ph{-}e^{2i\th_{r}\s^z}$	&	$\ph{-}\id$		\\	
10	&	$P_y^{-1}\T^{-1}P_y\T$			&	$\ph{-}e^{2i\th_{py}\s^z}$	&	$\ph{-}e^{2i\th_t\s^z}$
\\
11	&	$T_x^{-1}\T^{-1}T_x\T$			&	$-e^{2i\(\th_t+\th_{tx}\)\s^z}$	& $-\id$	\\
12	&	$T_y^{-1}\T^{-1}T_y\T$			&	$-e^{2i\(\th_t+\th_{ty}\)\s^z}$	&	$-\id$	\\
13	&	$\T^2$						&	$-\id$	&	$\ph{-}e^{2i\th_t\s^z}$	
\\\hlinewd{.8pt}
\end{tabular}}
\caption{Symmetry fractionalization of U(1) spin liquids.}
\label{tab:U1SymmFrac}
\end{table}

Wen \cite{Wen02} finds that the following U(1) phases are proximate to SU2B$n0$:
\begin{center}
\begin{tabular}{cccc}
U1B$000n$   & U1B$n10n$    & U1C$n0nn$  &   U1C$11nn$ \\
U1B$0001$   &   U1B$x10x$   &   U1C$n0n1$   &   U1C$11nx$   \\
U1B$001n$   &   U1B$x11n$   &   U1C$n0x1$   &   U1C$11xn$   \\
U1B$0011$   &   U1B$x11x$   &   U1C$n01n$   &   U1C$11xx$
\end{tabular}
\end{center}
In this section we determine which of these lattice PSG's corresponds to the gapped and gapless U(1) spin liquids obtained by condensing $\Phi$ and $\Phi_1$ respectively.

\subsubsection{Gapped U(1) spin liquid ($D_f$)}

To compare with Wen's classification, we condense the Higgs' fields in the $z$ component. 
Therefore, for the gapped U(1) spin liquid $D_f$, only $\Braket{\Phi^z}\neq0$ and the PSG in Eq.~\eqref{eqn:U1GappedPSGCont} should be rewritten:
\eq{
V_{tx}&=e^{i\th_{tx}\s^z}i\s^x,
&
V_{px}&=e^{i\th_{py}\s^z},
&
V_{r}&=e^{i\th_r\s^z},
\nt 
V_{ty}&=e^{i\th_{ty}\s^z}i\s^x,
&
V_{py}&=e^{i\th_{py}\s^z},
&
V_t&=e^{i\th_t\s^z}i\s^x,
}
The resulting symmetry fractionalization is shown in Table~\ref{tab:U1SymmFrac}.

We identify this phase in several steps.
We note that independent from $\th_t$, $(W_tU_t)^2=-\id$, and therefore, of the spin liquids proximate to SU2B$n0$, only those with $W_t(\vi)\not\propto \t^0$ are possible candidates.
Moreover, the U1B spin liquids all have $W_{tx}=(-)^{i_y}g_3(\th_{tx})$, $W_{ty}(\vi)=g_3(\th_{ty})$ where $g_\ell(\th)=e^{i\th \t^\ell}$.
Inserting these into group relation \#1 in Eq.~\eqref{eqn:LatticeGpRels} returns $-\id$, again independent of the angles $\th_{tx}$ and $\th_{ty}$, invalidating these options. 
This leaves four candidates: U1C$n0n1$, U1C$n0x1$, U1C$11nx$, and U1C$11xx$.
We have computed the symmetry fractionalization of each of these phases and determined that $D_f$ corresponds to U1C$n0n1$ whose lattice PSG is
\eq{\label{eqn:U1Cn0n1PSG}
\mathrm{U1C}n0n1&:
\nt
W_{tx}(\vi)&=(-)^{i_y}g_3(\th_{tx})i\t^1,
&
W_{ty}(\vi)&=g_3(\th_{ty})i\t^1,
\nt 
W_{px}(\vi)&=(-)^{i_x}g_3(\th_{px}),
&
W_{py}(\vi)&=(-)^{i_y}g_3(\th_{py}),
\nt 
W_{pxy}(\vi)&=(-)^{i_xi_y}g_3\(\th_{pxy}\),
&
W_r(\vi)&=(-)^{i_xi_y+i_x}g_3\(\th_r\)
\nt 
W_{t}(\vi)&=(-)^{i_x+i_y}g_3(\th_t)i\t^1.
}

\subsubsection{Gapless U(1) spin liquid}
The (continuum) PSG of the gapless spin liquid with $\Braket{\Phi_1^z}\neq0$ is 
\eq{
V_{tx}&=e^{i\th_{tx}\s^z},
&
V_{px}&=e^{i\th_{py}\s^z},
&
V_{r}&=e^{i\th_r\s^z},
\nt 
V_{ty}&=e^{i\th_{ty}\s^z}i\s^x,
&
V_{py}&=e^{i\th_{py}\s^z}i\s^x,
&
V_t&=e^{i\th_t\s^z}.
}
From the symmetry fractionalization in Table~\ref{tab:U1SymmFrac} and the arguments in the previous section, we conclude that only U1B spin liquids with $W_t\propto \t^0$ are possible candidates: U1B$000n$, U1B$n10n$, U1B001$n$, U1B$x11n$.
Computing the symmetry fractionalization of these four spin liquid identifies U1B$x11n$ as the correct lattice analogue:
\eq{\label{eqn:GapplessU1PSG}
\mathrm{U1B}x11n&:
\nt
W_{tx}(\vi)&=(-)^{i_y}g_3(\th_{tx})\t^0,
&
W_{ty}(\vi)&=g_3(\th_{ty})\t^0,
\nt 
W_{px}(\vi)&=(-)^{i_x}g_3(\th_{px})i\t^1,
&
W_{py}(\vi)&=(-)^{i_y}g_3(\th_{py})i\t^1,
\nt 
W_{pxy}(\vi)&=(-)^{i_xi_y}g_3(\th_{pxy})i\t^1,
&
W_r(\vi)&=(-)^{i_xi_y+i_x}g_3\(\th_r\)
\nt 
W_{t}(\vi)&=(-)^{i_x+i_y}g_3(\th_t)\t^0.
}
In Appendix~\ref{app:GaplessU1Ansatz} we show that this PSG has no lattice realization.

\subsection{$\Zt$ spin liquids}\label{app:Z2latticePSGcorr}

Wen divides the $\Zt$ spin liquids into two classes.
Their PSG's are
\eq{\label{eqn:Z2spinLiqDef}
W_{tx}(\vi)&=\tilde{\eta}^{i_y}\t^0,
&
W_{px}(\vi)&=\eta_{xpx}^{i_x}\eta_{xpy}^{i_y}g_{py},
&
W_{pxy}(\vi)&=(-)^{i_xi_y}g_{pxy},
\nt 
W_{ty}(\vi)&=\t^0,
&
W_{py}(\vi)&=\eta_{xpy}^{i_x}\eta_{xpx}^{i_y}g_{py},
&
W_t(\vi)&=\eta_t^{i_x+i_y}g_t,
}
where A spin liquids have $\tilde{\eta}=+1$ and B spin liquids have $\tilde{\eta}=-1$.
Unlike for the SU(2) case, each of the group elements $g_\xi$ takes only a single value.
He labels these spin liquids by $\mathrm{Z2A}\(g_{px}\)_{\eta_{xpx}}\(g_{py}\)_{\eta_{xpy}}g_{pxy}\(g_t\)_{\eta_t}$ and $\mathrm{Z2B}\(g_{px}\)_{\eta_{xpx}}\(g_{py}\)_{\eta_{xpy}}g_{pxy}\(g_t\)_{\eta_t}$.
An equivalent short-hand notation replaces $(\t^0,\t^1,\t^2,\t^3)$ and $(\t_+^0,\t_+^1,\t_+^2,\t_+^3)$ by $\(0,1,2,3\)$ and $(\t_-^0,\t_-^1,\t_-^2,\t_-^3)$ by $\(n,x,y,z\)$ (this is the notation used in the majority of the paper). 
There are 272 distinct such PSG's; however, though at least 72 of these are anomalous and cannot be described with a mean field Hamiltonian on the lattice. 

We can determine the symmetry fractionalization of each of these PSG's using Eq.~\eqref{eqn:LatticeGpRels}, forming a table similar to Table~\ref{tab:SymmFrac}, and this information is what leads to the identification in Table~\ref{tab:LatticeNames}.
It is clear that the symmetry fractionalization does not completely determine the PSG since both $s$PSG1 and $s$PSG5 have the same symmetry fractionalization as two different spin liquids. 
We will show that in both cases, a single lattice PSG can be associated with each of our continuum versions.

Our primary strategy will be to check that which PSG's in Table~\ref{tab:LatticeNames} are proximate to U1C$n0n1$. 
By studying Table~\ref{tab:U1SymmFrac}, we determine which values of $\th_\xi$ give the $\Zt$ symmetry fractionalization of the phases we're interested in.
In both cases we find only a single possibility.
We also verify that $s$PSG5 is proximate to U1B$x11n$.

We note that the symmetry transformations in Table~\ref{tab:U1SymmFrac} depend on only five generators: $T_x,T_y,P_y,R_{\pi/2},\T$.
To make contact with Wen's conventions, we also display the gauge transformations corresponding to $P_x=R_{\pi/2}P_yR_{\pi/2}^{-1}$ and $P_{xy}=R_{\pi/2}P_y^{-1}$; their forms are also determined by the angles $\th_{tx},\th_{ty},\th_{py},\th_{r},$ and $\th_{t}$.

\subsubsection{Lattice PSG of $A_f$ phase ($s$PSG5)}

We begin by determining which choice of angles of the gapped U(1) spin liquid returns the symmetry fractionalization of $s$PSG5.
Setting $\th_{tx}=0$ fixes the remaining angles to be
\eq{
\th_{ty}&=\pi,
&
\th_{py}&=\pm{\pi\o2},
&
\th_r&=0,\pi,
&
\th_t&=0,\pi.
}
The choices only result in gauge transformations differing by a minus sign and, except for $W_{ty}$, do not affect the symmetry fractionalization. 
In what follows we choose positive prefactors for all of the gauge transformations below.
Modulo these considerations, this is the \emph{only} PSG proximate to U1C$n0n1$ with the same symmetry fractionalization as $s$PSG1.
This gives
\eq{\label{eqn:sPSGgapped1}
W_{tx}&=(-)^{i_y}i\t^1,
&
W_{ty}&=-i\t^1,
\nt 
W_{px}&=(-)^{i_x}i\t^3,
&
W_{py}&=(-)^{i_y}i\t^3,
\nt 
W_{pxy}&=(-)^{i_xi_y}i\t^3,
&
W_{r}&=(-)^{i_xi_y+i_x}\t^0,
\nt 
W_t&=(-)^{i_x+i_y}i\t^1.
}
We can bring it into the form of Eq.~\eqref{eqn:Z2spinLiqDef} by performing the gauge transformation
\eq{\label{eqn:GaugeTransZ2}
W(\vi)&=\begin{cases}
(-)^{(i_x+i_y)/2}i\t^2,   &   i_x+i_y=even,
\\
(-)^{(i_x+i_y-1)/2}i\t^3    &   i_x+i_y=odd.
\end{cases}
}
Under this transformation, the PSG in Eq.~\eqref{eqn:sPSGgapped1} becomes
\eq{
W_{tx}&=(-)^{i_y}\t^0,
&
W_{ty}&=-\t^0,
\nt 
W_{px}&=(-)^{i_x+i_y}i\t^3,
& 
W_{py}&=(-)^{i_x+i_y}i\t^3,
\nt 
W_{pxy}&=(-)^{i_x(i_y+1)}i\t^3,
&
W_r&=(-)^{i_x(i_y+1)}\t^0,
\nt 
W_t&=(-)^{i_x+i_y}i\t^1.
}
Upon shifting $i_y\to i_y+1$, we recognize this PSG as $\mathrm{Z2B}zz3x$, and, rotating by 90$^0$ about the $y$-axis this becomes $\mathrm{Z2B}xx1z$.
This identifies $\mathrm{Z2B}xx1z$ as the unique lattice PSG capable of describing the phase $A_f$.

Another way we could have reached this conclusion is by studying the mean field ansatz allowed by either of these PSG's.
It turns out that the mean field Hamiltonian corresponding to the other candidate PSG, $\mathrm{Z2B}xx2z$, cannot be gapped, whereas no such restrictions exist for $\mathrm{Z2B}xx1z$.

We also show that $\mathrm{Z2B}xx1z$ is proximate to the gapless spin liquid $\mathrm{U1B}x11n$.
In order to reproduce the symmetry fractionalization of $s$PSG5, the angles in Eq.~\ref{eqn:GapplessU1PSG} must be
\eq{
\th_{tx}&=\pm{\pi\o2}
&
\th_{ty}&=\mp{\pi\o2},
&
\th_{r}&=0,\pi,
&
\th_{t}&=\pm{\pi\o2}.
}
$\th_{py}$ is un-determined, and therefore, unlike in the previous case, proximity to $\mathrm{U1B}x11n$ does not fully determine the lattice PSG corresponding to $s$PSG5.
The angles which are restricted indicate that
\eq{
W_{tx}&=(-)^{i_y}i\t^3,
&
W_{ty}&=-i\t^3,
\nt 
W_r&=(-)^{i_x(i_y+1)}\t^0,
& 
W_t&=(-)^{i_x+i_y}i\t^3.
}
Rotating by $90^0$ about the $y$-axis take $\t^3\to\t^1$.
We then observe that all of the gauge transformations shown above are equal to the corresponding gauge transformation in Eq.~\eqref{eqn:sPSGgapped1}.
It can be shown that $\th_{py}$ can be chosen to obtain $\mathrm{Z2B}xx1z$ but not $\mathrm{Z2B}xx2z$.
Therefore, only $\mathrm{Z2B}xx1z$ is proximate to $\mathrm{U1B}x11n$. .

\subsubsection{$s$PSG1}

Performing the same analysis as above, we find that the only way for the symmetry fractionalization of U1C$n0n1$ to return the symmetry fractionalization of $s$PSG1 is if the angles in Eq.~\eqref{eqn:U1Cn0n1PSG} are
\eq{
\th_{ty}&=\pi,
&
\th_{py}&=\pm{\pi\o2},
&
\th_{r}&=0,\pi,
&
\th_t&=\pm{\pi\o2},
}
where, again, we've set $\th_{tx}=0$. 
The gauge transformations associated with the symmetry generators are then
\eq{
W_{tx}&=(-)^{i_y}i\t^1,
&
W_{ty}&=-i\t^1,
\nt 
W_{px}&=(-)^{i_x}i\t^3,
&
W_{py}&=(-)^{i_y}i\t^3,
\nt 
W_{pxy}&=(-)^{i_xi_y}i\t^3,
&
W_{r}&=(-)^{i_x(i_y+1)}\t^0,
\nt 
W_t&=(-)^{i_x+i_y}i\t^2.
}
Performing the gauge transformation in Eq.~\eqref{eqn:GaugeTransZ2}, these become,
\eq{
W_{tx}&=(-)^{i_y}\t^0,
&
W_{ty}&=-\t^0,
\nt 
W_{px}&=i(-)^{i_x+i_y}i\t^3,
& 
W_{py}&=i(-)^{i_x+i_y}i\t^3,
\nt 
W_{pxy}&=(-)^{i_x(i_y+1)}i\t^3,
&
W_r&=(-)^{i_x(i_y+1)}\t^0,
\nt 
W_t&=i\t^2.
}
It is not difficult to see that this corresponds to $\mathrm{Z2B}zz32$, which is equivalent to $\mathrm{Z2B}xx13$.



\section{Lattice realizations of spin liquids}
\label{app:LatticeHamiltonians}
In this appendix, we use the lattice PSG's determined in Appendix~\ref{app:WenClassification} for the $\pi$-flux phase and $A_f$, $B_f$, and $D_f$ to write down the corresponding lattice Hamiltonian.
Doing so will serve as further verification of the symmetry fractionalization used in the main text.
Further, the calculation of the Berry phase in Sec.~\ref{sec:BerryPhaseSubSec} requires the lattice description of the gapped U(1) spin liquid corresponding to U1C$n0n1$.

\subsection{SUB$n0$ mean field Hamiltonian}\label{app:SU2Bn0latHam}
The ansatz for the $\pi$-flux state is given in Eq.~\eqref{eqn:SU2latPSG}.
Gauge invariance and the form of the translational symmetry operations compels the mean field parameters to take the following form:
\eq{
u_{\vi,\vi+\vm}=(-)^{i_xm_y}iu_\vm^0.
}
In order for the mean field Hamiltonian to be Hermition, $u^\dag_{\vi\vj}$ must equal $u_{\vj\vi}$.
This can be used to show that
\eq{
(-)^{i_xm_y}iu_\vm^0=-
(-)^{i_xm_y}(-)^{m_xm_y}iu_\vm^0,
}
which indicates
\eq{\label{eqn:SU2minusM}
u_{-\vm}^0&=-(-)^{m_xm_y}u^0_\vm.
}
Next, Eq.~\eqref{eqn:AnsatzInvariance} states that $u_{\vi\vj}$ must be invariant under the action of all (projective) symmetry operations.
In particular, acting $P_xP_y$ and using Eq.~\eqref{eqn:SU2minusM}, we find
\eq{
u_{\vi,\vi+\vm}&=W_{px}P_xW_{py}P_y[u_{\vi,\vi+\vm}]=-
(-)^{i_xm_y}(-)^{m_xm_y}(-)^{m_x+m_y}iu_\vm^0\t^0 .
}
Similarly, the action of time reversal requires
\eq{
u_{\vi,\vi+\vm}&=W_t\T[u_{\vi,\vi+\vm}]=
(-)^{i_xm_y}(-)^{m_x+m_y}iu_\vm^0\t^0.
}
Between these two equations, we conclude that $u_\vm^0\neq0$ only when $m_x+m_y=odd.$
Finally, we relate mean field parameters for different $\vm$'s through the action of $P_x$, $P_y$, and $P_{xy}$:
\eq{
u_{(-m_x,m_y)}^0&=(-)^{m_x}u_{(m_x,m_y)}^0,
&
u_{(m_x,-m_y)}^0&=(-)^{m_y}u_{(m_x,m_y)}^0,
&
u_{(m_y,m_x)}^0&=(-)^{m_xm_y}u_{(m_x,m_y)}^0.
}
The mean field ansatz we obtain is
\eq{\label{eqn:SU2nn}
u_{\vi,\vi+\hx}&=i\a\t^0,
&
u_{\vi,\vi+\hy}&=(-)^{i_x}i\a\t^0.
}

Inserting these hopping terms into Eq.~\eqref{eqn:MFgeneral} (and dropping the constant) we obtain
\eq{
H_{\pi}'&=-i\a\sum_\vi\( \psi^\dag_\vi\psi_{\vi+\hx}+(-)^{i_x}\psi_\vi^\dag\psi_{\vi+\hy}+h.c.\).
}
We now show that the low-energy theory is precisely the Dirac Hamiltonian.
In momentum space, we find
\eq{\label{eqn:HpiWen}
H_{\pi}'&=2\a\int_{-\pi/2}^{\pi/2}{dk_x\o2\pi}\int_{-\pi/2}^{\pi/2}{dk_y\o2\pi}\,
\Psi_\vk^\dag \(\sin k_x \tilde{\t}^3\m^3\t^0+\sin k_y\tilde{\t}^1\m^3\t^0\)
\Psi_\vk
}
where $\Psi_\vk=\(\psi_\vk,\psi_{\vk+\Q_x+\Q_y},\psi_{\vk+\Q_x},\psi_{\vk+\Q_y}\)^T$ with $\Q_x=(\pi,0)$ and $\Q_y=(0,\pi)$, and 
\eq{\label{eqn:matrixDefs}
\tilde{\t}^3\m^3&=\begin{pmatrix}
1   &   0   &   0   &   0 \\
0   &   -1   &   0   &   0 \\  
0   &   0   &   -1   &   0 \\ 
0   &   0   &   0   &   1
\end{pmatrix},
&
\tilde{\t}^1\m^3&=\begin{pmatrix}
0   &   0   &   1   &   0 \\ 
0   &   0   &   0   &   -1 \\ 
1   &   0   &   0   &   0 \\ 
0   &   -1   &   0   &   0 
\end{pmatrix},
&
\tilde{\t}^0\m^1&=\begin{pmatrix}
0   &   1   &   0   &   0 \\ 
1   &   0   &   0   &   0 \\ 
0   &   0   &   0   &   1 \\ 
0   &   0   &   1   &   0 
\end{pmatrix}.
}
Equivalently, writing $\Psi_\vk=\(\psi_{1,1,\vk},\psi_{1,2,\vk},\psi_{2,1,\vk},\psi_{2,2,\vk}\)^T$, we can identify the $\tilde{\t}^\ell$'s with Pauli matrices acting on the first index of $\Psi_\vk$ and the $\m^\ell$'s with Pauli matrices acting on the second.
Finally, to make contact with the expression in Sec.~\ref{sec:PiFluxAnsatz}, we express $H_\pi'$ in terms of 
\eq{\label{eqn:newPsi}
\tilde{\Psi}_\vk=e^{i\pi\tilde{\t}^2\m^3/4}\Psi_\vk.
}
The resulting mean field Hamiltonian is
\eq{
H_\pi'&=2\a\int_{-\pi/2}^{\pi/2}{dk_x\o2\pi}\int_{-\pi/2}^{\pi/2}{dk_y\o2\pi}\,
\tilde{\Psi}_\vk^\dag\(\sin k_x \t^1\m^0\s^0-\sin k_y\t^3\m^0\s^0\)
\nt&
\cong -2\a\int {d^2k\o(2\pi)^2}\,\tilde{\Psi}^\dag\( k_x \g^0\g^x + k_y\g^0\g^y \) \tilde{\Psi}_\vk
}
where we've rewritten the gauge-charged $\t^\ell$'s as $\s^\ell$'s (as done in the main body of the text) and used the fact that $\g^\m=(\tilde{\t}^y,i\tilde{\t}^z,i\tilde{\t}^x).$
It is clear that once dynamic gauge fields are included, this is equivalent to $\L_\qcd$ in Eq.~\eqref{eqn:LqcdDirac}.

\subsection{U1C$n0n1$ mean field Hamiltonian}\label{sec:GappedU1Ham}

We now use the ansatz for Eq.~\eqref{eqn:U1Cn0n1PSG} to determine the lattice Hamiltonian corresponding to the gapped spin liquid phase $D_f$. We show that it is precisely $H_\pi'$ plus a term which breaks the SU(2) symmetry to U(1): $H_{D_f}=H_\pi'+H_1$.

Eq.~\eqref{eqn:U1Cn0n1PSG} indicates that all bonds must be of the form
\eq{
u_{\vi,\vi+\vm}=(-)^{i_xm_y}\(iu_\vm^0\t^0+(-)^{i_x+i_y}u^3_\vm \t^3\).
}
Further, hermiticity of the Hamiltonian requires $u_{\vi\vj}^\dag = u_{\vj\vi}$ and therefore
\eq{
(-)^{i_xm_y}\(-iu_\vm^0\t^0+(-)^{i_x+i_y}u_\vm^3\t^3\)
=
(-)^{i_xm_y}(-)^{m_xm_y}\(iu_{-\vm}^0\t^0+(-)^{i_x+i_y}(-)^{m_x+m_y}u_{-\vm}^3\t^3\)
,}
implying that 
\eq{
u_\vm^0&=-(-)^{m_xm_y}u_{-\vm}^0,
&
u_{\vm}^3&=(-1)^{m_x+m_y}(-)^{m_xm_y}u_{-\vm}^3 \,.
}
Similarly, to satisfy Eq.~\eqref{eqn:AnsatzInvariance}, $u_{\vi\vj}$ must be invariant under 180$^0$ rotations:
\eq{
u_{\vi,\vi+\vm}=
W_{px}P_xW_{py}P_y[u_{\vi,\vi+\vm}]&=(-)^{i_xm_y}(-)^{m_xm_y}\Big[-(-)^{m_x+m_y}iu_{\vm}^0\t^0+(-)^{i_x+i_y}u^3_{\vm}\t^3\Big],
}
where we've used the previous expression to relate $u_{\vm}^\ell$ and $u_{-\vm}^\ell$. 
It follows that $u_\vm^0=0$ when $(m_x,m_y)=(even,even)$ and that $u_\vm^3=0$ when $(m_x,m_y)=(odd,odd).$ 
The ansatz must also be invariant under $\T$:
\eq{
u_{\vi,\vi+\vm}&=W_t\T[u_{\vi,\vi+\vm }]=(-)^{i_xm_y}(-)^{m_x+m_y}\(-iu_{\vm}^0\t^0+(-)^{i_x+i_y}u^3_\vm\t^3\),
}
showing that $u^0_\vm$ is non-zero only for $m_x+m_y=odd$ and that $u_\vm^3$ is only non-zero when $m_x+m_y=even$.
Together, these give
\eq{
u_{\vi,\vi+\vm}
&=
\begin{cases}
(-)^{i_x+i_y}u_\vm^3,
&
(m_x,m_y)=(even,even),
\\
(-)^{i_xm_y}iu_\vm^0\t^0,
&
m_x+m_y=odd.
\end{cases}
}
We can also show that the action of $P_x$, $P_y$, and $P_{xy}$ implies the following relations:
\eq{
u^\ell_{(m_x,my)}&=(-)^{m_x}u^\ell_{(-m_x,m_y)},
&
u^\ell_{(m_x,my)}&=(-)^{m_y}u^\ell_{(m_x,-m_y)},
&
u^\ell_{(m_x,m_y)}&=(-)^{m_xm_y}u^\ell_{(m_y,m_x)},
}
for $\ell=0,3$.
Using these relations, we find,
\eq{\label{eqn:GappedU1Ansatz}
u_{\vi,\vi+\hx}&=i\a \t^0,
&
u_{\vi,\vi+2\hx}&=(-)^{i_x+i_y}\b\t^3,
&
u_{\vi,\vi}&=(-)^{i_x+i_y}a_0\t^3,
\nt 
u_{\vi,\vi+\hy}&=(-)^{i_x}i\a \t^0,
&
u_{\vi,\vi+2\hx}&=(-)^{i_x+i_y}\b\t^3.
}
As expected, the nearest-neighbour bonds are identical to those we found for the $\pi$-flux phase in the previous section.
The SU(2) symmetry is already broken to U(1) by the inclusion of the next-nearest neighbour bonds and so this is all we consider.

As in the previous section, the mean field Hamiltonian is obtained by inserting these hopping terms into Eq.~\eqref{eqn:MFgeneral}:
\eq{
H_{D_f}&=H_\pi'+H_1,
\nt
H_1&=\sum_\vi(-)^{i_x+i_y}\bigg[
\b\(\psi_\vi^\dag \t^3\psi_{\vi+2\hx}+\psi_\vi^\dag\t^3\psi_{\vi+2\hy}+h.c.\)
-a_0^3\psi^\dag_\vi\t^3\psi_\vi
\bigg], \label{U1MF}
}
where $H_\pi'$ is given above in Eq.~\eqref{eqn:HpiWen}.
In momentum space, this becomes
\eq{
H_{D_f}&=\int_{-\pi/2}^{\pi/2}{dk_x\o2\pi}\int_{-\pi/2}^{\pi/2}{dk_y\o2\pi}\,
\Psi_\vk^\dag \bigg[
2\a\(\sin k_x \tilde{\t}^3\m^3\t^0+\sin k_y\tilde{\t}^1\m^3\t^0\)
\nt&\quad-\(2\b \[\cos 2k_x +\cos2k_y\]-a_0\)\tilde{\t}^0\m^1\t^3 \bigg]\Psi_\vk
}
where we've used the same notation as in the previous section: $\Psi_\vk=\(\psi_\vk,\psi_{\vk+\Q_x+\Q_y},\psi_{\vk+\Q_x},\psi_{\vk+\Q_y}\)^T$ with $\Q_x=(\pi,0)$ and $\Q_y=(0,\pi)$, and the matrices defined in Eq.~\eqref{eqn:matrixDefs}.
In terms of $\tilde{\Psi}_\vk=e^{i\pi\tilde{\t}^2\m^3/4}\Psi_\vk$:
\eq{
H_{D_f}&=\int_{-\pi/2}^{\pi/2}{dk_x\o2\pi}\int_{-\pi/2}^{\pi/2}{dk_y\o2\pi}\,
\tilde{\Psi}_\vk^\dag \bigg[
2\a\(\sin k_x \tilde{\t}^1\m^0\s^0-\sin k_y\tilde{\t}^3\m^0\s^0\)
\nt&\quad-\(2\b \[\cos 2k_x +\cos2k_y\]-a_0\)\tilde{\t}^2\m^2\s^3 \bigg]\tilde{\Psi}_\vk,
}
where, again, we've rewritten the SU(2) matrices $\t^\ell$ as $\s^\ell$ in accord with the continuum notation.
Expanding $H_{D_f}$ about $\vk=(0,0)$, we obtain
\eq{
H_{D_f}\cong \int {d^2k\o(2\pi)^2}\,\tilde{\Psi}^\dag\bigg[
-2\a\( k_x \g^0\g^x + k_y\g^0\g^y \) 
+\(4\b-a_0\)\g^0\m^y\s^z \Big]\tilde{\Psi}_\vk
}
where $\g^\m=(\tilde{\t}^y,i\tilde{\t}^z,i\tilde{\t}^x).$
We conclude that the term which reduces the SU(2) symmetry down to U(1) is precisely equivalent to $\bpsi \m^y\s^z\psi \sim \tr\(\s^z\bX \m^y X\)$.


\subsection{U1B$x11n$ mean field Hamiltonian}\label{app:GaplessU1Ansatz}

In this subsection, we demonstrate that U1B$x11n$ has \emph{no} lattice analogue.
Referring to Eq.~\ref{eqn:GapplessU1PSG}, we see that gauge and translational symmetry requires
\eq{
u_{\vi,\vi+\vm}=(-)^{i_xm_y}\(iu_\vm^0\t^0 +u_\vm^3 \t^3\).
}
We relate $u_{-\vm}^{0,3}=u_{\vm}^{0,3}$ using the fact that $u_{\vi,\vi+\vm}^\dag=u_{\vi+\vm,\vi}$:
\eq{
u_{-\vm}^0&=-(-)^{m_xm_y}u_\vm^0,
&
u_{-\vm}^3&=(-)^{m_xm_y}u_\vm^1.
}
Then, acting on $u_{\vi,\vi+\vm}$ with $P_xP_y$ and $\T$ gives
\eq{
W_{px}P_xW_{py}P_y[u_{\vi,\vi+\vm}]&=(-)^{i_xm_y}(-)^{m_xm_y}(-)^{m_x+m_y}\(-iu_\vm^0\t^0+u_\vm^3\t^3\),
\nt
\T[u_{\vi,\vi+\vm}]&=
(-)^{i_xm_y}(-)^{m_x+m_y}\(-iu_\vm^0\t^0-u_\vm^3\t^3\).
}
Equating these expressions with $u_{\vi,\vi+\vm}$ implies that $u_\vm^0\neq0$ only for $m_x+m_y=odd$, as for SU2B$n0$ and U1C$n0n1$; it can be shown that they must satisfy identical constraints as the $\t^0$-bonds allowed by these PSG's. 
In particular, the nearest-neighbour values are identical to those in Eq.~\eqref{eqn:SU2nn}.
Conversely, there are no consistent solutions for $u_\vm^3$: it always vanishes and is therefore unable to break the SU(2) gauge symmetry to U(1).

\subsection{Z2B$xx1z$ mean field Hamiltonian}\label{app:Z2Ham}
We choose a gauge such that Eq.~\eqref{eqn:sPSGgapped1} describes the PSG of $\mathrm{Z2B}xx1z$. Translational symmetry and gauge invariance implies that
\eq{
u_{\vi,\vi+\vm}&=(-)^{i_xm_y}\(iu_\vm^0\t^0 + u_\vm^1 \t^1 + (-)^{i_x+i_y}\[ u_\vm^2\t^2+ u_\vm^3\t^3\]\).
}
Hermiticity then requires
\eq{
u_{-\vm}^0&=-(-)^{m_xm_y}u_\vm^0,
&
u_{-\vm}^1&=(-)^{m_xm_y} u_\vm^1,
&
u_{-\vm}^{2,3}&=(-)^{m_xm_y}(-)^{m_x+m_y}u_\vm^3.
}
Under the action of $P_xP_y$ and $\T$ the ansatz transforms as
\eq{
W_{px}P_xW_{py}P_y[u_{\vi,\vi+\vm}]&=(-)^{i_xm_y}(-)^{m_xm_y}\Big(-(-)^{m_x+m_y}iu_\vm^0\t^0+(-)^{m_x+m_y}u_\vm^2\t^1
\nt&\quad
+(-)^{i_x+i_y}\[u_\vm^2\t^2+u_\vm^3\t^3\]\Big),
\nt
\T[u_{\vi,\vi+\vm}]&=
(-)^{i_xm_y}(-)^{m_x+m_y}\(-iu_\vm^0\t^0-u_\vm^1\t^1+(-)^{i_x+i_y}\[ u_\vm^2\t^2 +u_\vm^3\t^3\]\).
}
These relations imply that $u_\vm^1=0$ for all $\vm$, $u_\vm^0\neq0$ only for $m_x+m_y=0$, and that $u_\vm^{2,3}\neq0$ only for $(m_x,m_y)=(even,even)$.
By studying the action of $P_x$, $P_y$, and $P_{xy}$, we obtain the following relations:
\eq{
u^0_{(-m_x,m_y)}&=(-)^{m_x}u^0_{(m_x,m_y)},
&
u^2_{(-m_x,m_y)}&=-u^2_{(m_x,m_y)},
&
u^3_{(-m_x,m_y)}&=u^3_{(m_x,m_y)},
\nt
u^0_{(m_x,-m_y)}&=(-)^{m_y}u^0_{(m_x,m_y)},
&
u^2_{(m_x,-m_y)}&=-u^2_{(m_x,m_y)},
&
u^3_{(m_x,-m_y)}&=u^3_{(m_x,m_y)},
\nt
u^0_{(m_y,m_x)}&=u^0_{(m_x,m_y)},
&
u^2_{(m_y,m_x)}&=-u^2_{(m_x,m_y)},
&
u^3_{(m_y,m_x)}&=u^3_{(m_x,m_y)}.
}
These show that $u^{0,3}_\vm$ are restricted to take the same values as in Eq.~\eqref{eqn:GappedU1Ansatz} for the gapped U(1) spin liquid, leaving the $u^2_\vm$ bonds to break the U(1) gauge symmetry down to $\Zt$. 
It turns out that its first non-zero value occurs at sixth nearest-neighbour:
\eq{
u_{\vi,\vi+2\hx+4\hy}&=(-)^{i_x+i_y}\g \t^2,
&
u_{\vi,\vi+2\hx-4\hy}&=-(-)^{i_x+i_y}\g \t^2,
\nt
u_{\vi,\vi+4\hx+2\hy}&=-(-)^{i_x+i_y}\g \t^2,
&
u_{\vi,\vi+4\hx-2\hy}&=(-)^{i_x+i_y}\g \t^2.
}
The contribution of these bonds to the Hamiltonian is 
\eq{
H_2=\g\sum_\vi (-)^{i_x+i_y}\bigg[
\psi_\vi^\dag \t^2 \psi_{\vi+2\hx+4\hy} - \psi_\vi^\dag \t^2 \psi_{\vi+2\hx-4\hy}-\psi_{\vi}^\dag \t^2 \psi_{\vi+4\hx+2\hy} 
+\psi_\vi^\dag \t^2 \psi_{\vi+4\hx-2\hy} -h.c. \bigg],
}
and the minimal Hamiltonian needed to describe Z2B$xx1z$ is $H_{A_f}=H_{D_f}+H_2$.
In momentum space, we have
\eq{
H_2&=4\gamma\int_{-\pi/2}^{\pi/2}{dk_x\o2\pi}\int_{-\pi/2}^{\pi/2}{dk_y\o2\pi}\,\(\sin4k_x\sin2k_y-\sin2k_x\sin4k_y\)\Psi_\vk^\dag \tilde{\t}^0\m^2\t^2\Psi_\vk,
}
where $\Psi_\vk$ in defined in Eq.~\eqref{eqn:newPsi},  and the action of the Pauli matrices $\tilde{\t}^\ell$ and $\m^\ell$ is given in Eq.~\eqref{eqn:matrixDefs} and below. 
Once more, we change notation such that Pauli matrices acting on colour space, $\t^\ell$, becomes $\s^\ell$'s and express $H_2$ in terms of the transformed fermion operator, $\tilde{\Psi}_\vk=e^{i\pi\tilde{\t}^2\m^3/4}\Psi_\vk$:
\eq{
H_2&=4\gamma\int_{-\pi/2}^{\pi/2}{dk_x\o2\pi}\int_{-\pi/2}^{\pi/2}{dk_y\o2\pi}\,\(\sin4k_x\sin2k_y-\sin2k_x\sin4k_y\)\tilde{\Psi}_\vk^\dag \t^y\m^y\s^y\tilde{\Psi}_\vk,
\nt
&\cong 
-16\gamma\int{d^2k\o(2\pi)^2}\,\tilde{\Psi}_\vk^\dag\Big[ k_xk_y\(k_x^2-k_y^2\)\g^0\m^y\s^y\Big]\tilde{\Psi}_\vk.
}
Notably, $H_2$ does \emph{not} correspond to any of the continuum operators in the action we study in the main text, in particular $\tr\(\s^a\bX \ptl_0 X\)\sim \bpsi \s^a\ptl_0\psi$. 
Instead, in the continuum language,
$H_2$ is proportional to $\bpsi \m^y\s^y\ptl_x\ptl_y\(\ptl_x^2-\ptl_y^2\)\psi\sim\tr\(\s^y\bX \m^y\ptl_x\ptl_y\[\ptl_x^2-\ptl_y^2\]X\)$.
This is discussed in Sec.~\ref{sec:PSGcorres}.

\section{Symmetry fractionalization of current-loop ordered spin liquid}\label{app:CLsymFrac}
\begin{table}
\centering
{\footnotesize
\begin{tabular}{|r| l | | r | r | r | r |}
\hline
& \multicolumn{1}{c||}{Group relations}	& fermionic & vison &   twist   & bosonic 
\\\hline\hline
1	&	$T_y^{-1}T_xT_yT_x^{-1}$	&   $-1$    &	$-1$    &   $1$    & $1$ 
\\\hline
2	&	$P_y^{-1}T_xP_yT_x^{-1}$	&	$-1$    &   $-1$    &   $1$    & $1$ 
\\\hline
3	&	$P_y^{-1}T_yP_yT_y$	&   $-1$    &   $1$    &    $1$    &    $-1$ 
\\\hline
4	&	$P_y^2$	&   $-1$    &   $1$    &    $-1$    &   $1$ 
\\\hline
5	&	$T_x^{-1}(\T P_x)^{-1}T_x(\T P_x)$		&   $-1$    &   $1$    & $1$    &    $-1$ 
\\\hline
6	&	$T_y^{-1}(\T P_x)^{-1}T_y(\T P_x)$	&   $1$    &    $-1$    &   $1$    &  $-1$ 
\\\hline
7	&	$P_y^{-1}(\T P_x)^{-1}P_y(\T P_x)$	&   $-1$    &   $-1$    &   $1$    &  $1$ 
\\\hline
8	&	$(\T P_x)^{2}$	&   $-1$    &   $1$    &    $1$    &    $-1$ 
\\\hline
\end{tabular}}
\caption{Symmetry fractionalization and twist factors for the fermionic and bosonic spinon and the vison in the phase $\Zt$ spin liquid with current-loop order. 
By comparing with the result in Ref.~\onlinecite{CSS17}, are able to verify the equivalent of $C_f$ and $C_b$.}
\label{tab:clPSGsymFrac}
\end{table}

We can also use symmetry fractionalization to verify that the phase $C_f$ corresponds to $C_b$.
There are now only eight group relations, and these are listed in Table~\ref{tab:clPSGsymFrac}.
The PSG of the reduced symmetry group is defined by the gauge transformations
\eq{
V_{tx}&=i\s^y,  
&
V_{py}&=i\s^x,
\nt 
V_{tx}&=i\s^y,
&
V_{tpx}&=i\s^z,
}
where the subscript $tpx$ denotes the joint group action of $\T P_x$.
With these, we determine the fermionic symmetry fractionalization using the methods described in Sec.~\ref{sec:SymFrac}. 
The results are shown in Table~\ref{tab:clPSGsymFrac} under the column labeled ``fermionic."

Both the symmetry fractionalization of the vison and the twist factors for the reduced symmetry relations can be worked out from the ones already given; Table~\ref{tab:clPSGsymFrac} lists these under the columns ``vison" and ``twist" respectively.

In order to determine the bosonic symmetry fractionalization, we borrow notation from Ref.~\onlinecite{CSS17}.
In Table~\ref{tab:BosonicSymTrans}, the symmetry transformation properties of the bosonic spinon and Higgs fields in Eqs.~\eqref{b1} and~\eqref{b2} are reproduced.
It will be convenient to express the bosonic spinon in terms of the four-component field $\mathcal{Z}=\(z,z^*\)^T=\(z_\uparrow,z_\downarrow,z^*_\uparrow,z_\downarrow^*\)^T$.
We then let Pauli matrices $\t^\ell$ act on this new index, while $\sigma$-matrices will act on the spin indices as before.
The U(1) gauge transformations are expressed as $\mathrm{U}(1)_g:\mathcal{Z}\to e^{i\th \t^z}\mathcal{Z}$.
In this language, the symmetry transformations are expressed as
\eq{
\T[\mathcal{Z}]&=i\s^y\t^z \mathcal{Z},
&
P_{x,y}[\mathcal{Z}]&=\mathcal{Z},
&
T_{x,y}[\mathcal{Z}]&=i\s^y\t^x \mathcal{Z}.
}
Using these, we obtain the numbers in the column of Table~\ref{tab:clPSGsymFrac} labeled ``boson." 

Finally, we multiply the twist, vison, and boson columns and obtain the numbers in the fermion column, thereby verifying the equivalence of $C_f$ and $C_b$.

\begin{table}
\centering
\begin{tabular}{| c || c | c | c | c | c |}
\hlinewd{.8pt}
    &   $\T$    & $P_x$ &   $P_y$   &   $T_x$   &   $T_y$ \\\hlinewd{.8pt}\hline 
$z_\a$    &   $i\s^y z$ &   $z$ &   $z$  &   $i\s^y z^*$ &  $i\s^y z^*$
\\\hline 
$Q_x$   &   $Q_x$   &   $-Q_x$ &   $Q_x$    &   $Q_x^*$ &   $Q_x^*$ 
\\\hline 
$P$ &   $-P$   & $P$ &   $P$ &   $P^*$   &   $P^*$
\\\hlinewd{.8pt}
\end{tabular}
\caption{Symmetry action on the bosonic spinon and Higgs fields in the bosonic dual to the theories studied here, as presented in Eq.~\eqref{b1} and Eq.~\eqref{b2} \cite{CSS17}. 
The spinon here is written as a two-component spinor, $z=\(z_\uparrow,z_\downarrow\)^T$ and that $i\s^y$ acts on these indices. We note that $\T[z^*]=-i\s^yz^*$ and that $T_{x,y}[z^*]=i\s^y z$.}
\label{tab:BosonicSymTrans}
\end{table}

\section{Linear response to nontrivial flux}
\label{app:fluxResponse}

\begin{figure}
    \centering
    \includegraphics[scale=1]{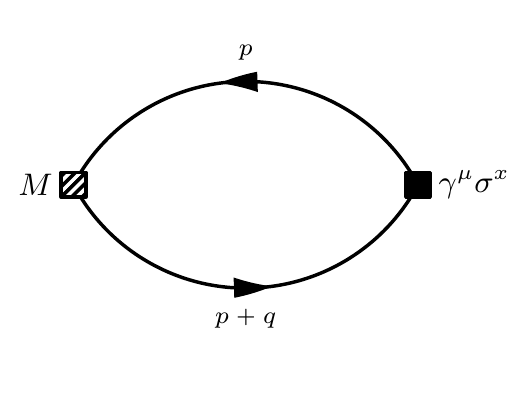}
    \caption{Fermion bubble to calculate flux response.}
    \label{fig:fluxResp}
\end{figure}

In this appendix, we calculate the relation in Eq.~\eqref{eqn:fluxLinResp} in imaginary time.
The residual U(1) gauge field $a_\m$ couples to the current $J^\m=\bpsi\g^\m\s^x\psi$.
The response function of an operator $\O$ is  $\chi_\O^\m=\Braket{\O(x)J^\m(x')}$, and the linear response equation in momentum space is simply $\Braket{\O(q)}=\chi_\O^\m(p)A_\m(q)$ (we specify to operators whose vacuum expectation values vanish in the absence of perturbations).
Assuming $\O=\tr\(\bX M X\)$, $\chi_0^\m(q)$ is represented by the Feynman diagram in Fig.~\ref{fig:fluxResp} at leading order.
We evaluate this as
\eq{
\chi_\O^\m(q)&=-\int{d^3p\o(2\pi)^3}\tr\bigg[ M{\sd{p}+im\s^x\m^y\o p^2+m^2}\g^\m\s^x {\sd{p}+\sd{q}+im\s^x\m^y\o(p+q)^2+m^2}\bigg].
}
If $M\propto \s^x$, it can be shown that the leading order term is quadratic in $q$. 
A $q$-linear piece is obtained by assuming that $\tr\(M\s^x\)=0$, in which case
\eq{
\chi_\O^\m(q)&=-m\int {d^3p\o(2\pi)^3}{1\o \[p^2+m^2\]\[(p+q)^2+m^2\]}\Big\{
p_\a\tr\[M\g^\a\g^\m\m^y\]
+(p+q)_\a\tr\[M\m^y\g^\m\g^\a\]
\Big\}
\nt
&=
{m\o8\pi}{iq_\a\o\abs{q}}\arctan\(\abs{q}\o2m\)A_\m(q)\(\tr\[\m^yM\g^\a\g^\m\]-\tr\[M\m^y\g^\m\g^\a\]\).
}
This is only non-zero for $M=\m^y\g^\n$.
Expanding the inverse tangent in small $q$, we find
\eq{
\chi_\n^\m(q)\cong - {1\o\pi}\ep^{\n\a\m}q_\a A_\m(q)\cong {i\o\pi} \ep^{\m\n\a}\ptl_\a A_\b(q).
}
Returning to real time, we obtain the result in Eq.~\eqref{eqn:fluxLinResp}.

\bibliography{higgs}
\end{document}